\documentclass[aps,prb,twocolumn,amsmath,amssymb,superscriptaddress,scrartcl,eqsecnum,longbibliography]{revtex4-1}
\usepackage{extarrows}
\usepackage{slashed}
\usepackage{amsmath,amsfonts,amssymb,graphics,graphicx,epsfig,color,times,indentfirst,layout}
\usepackage{lipsum}
\usepackage{mathrsfs}
\usepackage[unicode=true,pdfusetitle,bookmarks=false,colorlinks=true,citecolor=black,urlcolor=black,linkcolor=black]{hyperref}

\def\be{\begin{equation}}
\def\ee{\end{equation}}

\begin{document}

\rightline{MIT-CTP/4999}

\title{
Floquet conformal field theory
}

\date{\today}

\author{Xueda Wen}
\email{wenxueda@mit.edu}
\affiliation{Department of Physics, Massachusetts Institute of Technology, Cambridge, MA 02139, USA}

\author{Jie-Qiang Wu}
\affiliation{Center for Theoretical Physics, Massachusetts Institute of Technology, Cambridge MA, 02138 USA}

\begin{abstract}
Given a \textit{generic} two-dimensional conformal field theory (CFT),
we propose an analytically solvable setup to study the Floquet dynamics of the CFT,
\textit{i.e.}, the dynamics of a CFT subject to a periodic driving.
A complete phase diagram in the parameter space
can be analytically obtained within our setup.
We find two phases: the heating phase and the non-heating phase.
In the heating phase, the entanglement entropy keeps growing linearly in time, indicating that
the system keeps absorbing energy;
in the non-heating phase, the entanglement entropy oscillates periodically in time,
\textit{i.e.}, the system is not heated.
At the phase transition, the entanglement entropy grows logarithmically in time in a universal way.
Furthermore, we can obtain the critical exponent by studying
the entanglement evolution near the phase transition.
Mathematically, different phases (and phase transition) in a Floquet CFT
correspond to different types of M$\ddot{\text{o}}$bius transformations.
\end{abstract}
\maketitle

\textbf{\textit{Introduction}}$\,\,$
The dynamics of periodically driven (Floquet) many-body systems
has received extensive attentions recently. It sheds light on
fundamental issues in condensed matter physics and statistical physics such as the
phase structures and thermalization. Striking examples include Floquet topological insulators,\cite{Takashi2009,Takuya2010,lindner2011floquet,rechtsman2013photonic,cayssol2013floquet,
PhysRevX.3.031005,Titum2015,PhysRevX.6.021013,Jelena2016,Jelena2017}
Floquet symmetry protected/enriched topological phases,
\cite{Iadecola2015,Von2016,Else2016prb,Potter2016,PhysRevB.96.245116}
Floquet time crystals,
\cite{Else2016,Sondhi2016,Sondhi2016PRB,Else2017,Zhang2017,PhysRevLett.118.030401}
and Floquet thermodynamics.\cite{Moessner2014,Abanin2017,Abanin2015prl,kuwahara2016floquet,gritsev2017integrable}

In this work, we are interested in the Floquet dynamics of a (1+1) dimensional quantum critical point which
is described by a conformal field theory (CFT). To our knowledge, little attention has been paid in this direction.
In Ref.\onlinecite{Berdanier2017}, the Floquet dynamics of a \textit{boundary} driven quantum critical point was studied.
It was found that, depending on the driving frequency,
there are multiple dynamics regimes, including a heating regime and several other non-heating regimes.
Since the energy injected (from the boundary) per cycle is not
extensive in system size, it is still an open question on the Floquet dynamics of a bulk-driven quantum critical point.
It is well known that  CFTs after a quantum quench have brought to us much insight
in the non-equilibrium dynamics of many-body systems.\cite{CC2016,CC_Global,CC_Local}
Now, for a periodically bulk-driven CFT,
it is desirable to understand its Floquet dynamics. However, an analytically solvable setup is still lacking.

We fill this gap by proposing an analytically solvable setup for a bulk-driven Floquet CFT.
Both the correlation functions and the entanglement entropy can be analytically obtained in the whole parameter space
within our setup. We find two different phases depending on the driving frequency,
namely the heating and non-heating phases.
\footnote{
It is emphasized that here `heating' does not mean `thermalization'. As discussed in the following,
`heating' simply means that the system keeps absorbing energy in a quantum field theory with
infinite degrees of freedom.
}
In the heating phase, the entanglement entropy keeps growing linearly in time, which indicates that the system
keeps absorbing energy; in the non-heating phase, the entanglement entropy keeps oscillating in time,
indicating that the system is not heated. In particular, in the high frequency driving regime of the non-heating phase,
the oscillation period of entanglement entropy is independent of the driving frequency. In addition,
as we approach the phase transition, the oscillation period of entanglement entropy diverges, based on
which we can extract the critical exponent $\zeta=1/2$. The same critical exponent can be obtained
if we approach the phase transition from the heating phase, by studying the slope of the
linear growth of entanglement entropy.
At the phase transition, in the long time limit, the entanglement entropy grows logarithmically in time as $\frac{c}{3}\log t$,
where $c$ is the central charge of CFT.
We confirm our CFT result with a numerical simulation based on a free fermion lattice model.
We also find an elegant mathematical
structure underlying the phase diagram. The heating phase, non-heating phase and phase transition
in the Floquet CFT correspond to three kinds of M$\ddot{\text{o}}$bius transformations,
\textit{i.e.}, hyperbolic, elliptic, and parabolic
transformations, respectively. Our setup applies to a family of periodically driven CFTs.

\textbf{\textit{Our setup}}$\,\,$
Now we consider a generic (1+1) dimensional CFT defined on
a finite space of length $L$, with conformally invariant boundary conditions imposed at $x=0$ and $x=L$, respectively.
\footnote{We can also consider a system with periodic boundary condition if the Hamiltonian is composed of
three generators of Virasoro algebra $L_0$, $L_n$ and $L_{-n}$ with $n>1$.}
The initial state is prepared
as the ground state $|G\rangle$ of Hamiltonian $H_0$, and then we drive the system in the following way
\begin{equation}
\begin{gathered}
\includegraphics[width=7cm]{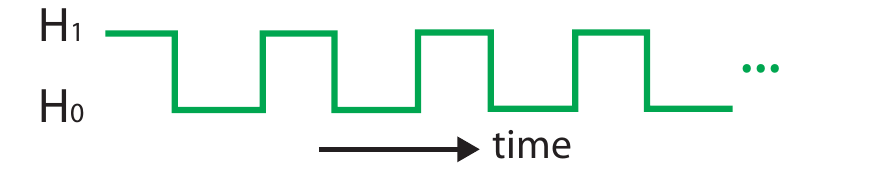}
\end{gathered}\nonumber
\end{equation}
Here $H_0$ denotes a uniform Hamiltonian of the form
\be\label{H0_maintext}
H_0=\int_0^L \frac{dx}{2\pi}T_{\tau\tau}(x),
\ee
where $T_{\tau\tau}(x)$ is the `time-time' component of the stress tensor.
For later convenience, we have defined our theory in Euclidean space with $w=\tau+ix$, so that
$T_{\tau\tau}=T_{ww}+T_{\bar{w}\bar{w}}=:T+\overline{T}$.
The other Hamiltonian $H_1$, among a family of candidates,\cite{WenUN}
is constructed by deforming $H_0$ as follows
\be\label{SSD}
H_1=H_0-\frac{1}{2}\left(H_++H_-\right),
\ee
where $H_{\pm}=\int_0^L\frac{dx}{2\pi}
\left(e^{\pm 2\pi w/L}T(w)+e^{\mp2\pi\bar{w}/L}\overline{T}(\bar{w})\right)$.
$H_1$ itself describes a sine-square deformed CFT which was extensively studied recently.
\cite{Gendiar2009,Gendiar2010,Hikihara2011,Gendiar2011,
Shibata2011,Shibata2012,Shibata2013,Katsura2011a,
Katsura1110,Maruyama2011,
Tada1404,Okunishi2015,Tada1504,Tada1602,Okunishi1603,Ryu1604,Katsura1709,Tada1712,WenWu2018}
It was found that a CFT with Hamiltonian $H_1$ has a continuous Virasoro
algebra that results in a continuous energy spectrum,\cite{Tada1504,Tada1602} which is in contrast with $H_0$
that has discrete energy spacing $\propto 1/L$.
\footnote{It is noted that the feature of continuous spectrum for $H_1$ is not essential here.
We can also consider the Hamiltonian $H_1=H_0-\frac{\tanh(2\theta)}{2}\left(H_++H_-\right)$ with $\theta\ge 0$.
For finite $\theta$, $H_1$ has discrete energy spacing $\propto 1/L\cosh(2\theta)$.
One can find similar physics in the Floquet CFT in this case.\cite{WenUN}
}
In short, starting from the ground state $|G\rangle$ of $H_0$, we drive
the system with $H_1$ for a time interval $T_1$, and then with $H_0$ for a time interval $T_0$, and repeat
this driving procedure. To characterize the Floquet dynamics of the system,
we study the correlation functions and entanglement entropy
evolution at time $t=n(T_0+T_1)$, with $n=0,1,2,\cdots$.

\begin{figure}[tp]
\includegraphics[width=3.5in]{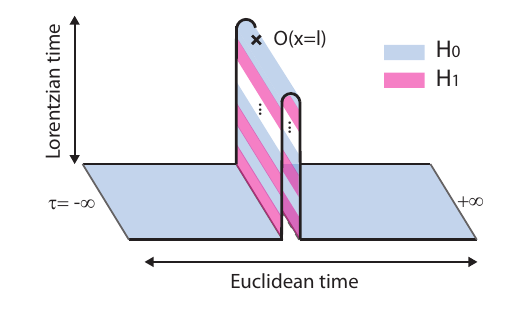}
\caption{Path integral representation of the single-point correlation function $\langle \psi(t)|\mathcal{O}(x)|\psi(t)\rangle$
in $w$-plane, with $w=\tau+ix$.
}
\label{FloquetPI}
\end{figure}

The wavefunction after $n$ cycles of driving can be written as
$
|\psi(t)\rangle=e^{-iH_0T_0}e^{-iH_1T_1}\cdots
e^{-iH_0T_0}e^{-iH_1T_1}
|G\rangle,
$
based on which we can evaluate the multi-point correlation functions.
For simplicity, now let us consider the single point correlation function $\langle \psi(t)|\mathcal{O}|\psi(t)\rangle$.
Its path integral representation in $w$-plane is shown in Fig.\ref{FloquetPI}, with $\tau\in (-\infty,\infty)$ and
$x\in[0,L]$, \textit{i.e.}, the path integral is defined on a strip.
Note there is Lorentz (real) time evolution introduced by the driving.
To evaluate $\langle \mathcal{O}\rangle$,
we go to the Euclidean space by writing $|\psi(t)\rangle$ as
$|\psi(\tau)\rangle=e^{-H_0\tau_0}e^{-H_1\tau_1}\cdots
e^{-H_0\tau_0}e^{-H_1\tau_1}|G\rangle$, and do the analytical continuation
$\tau_0\to iT_0$,  $\tau_1\to iT_1$ in the final step.

\begin{figure}[tp]
\includegraphics[width=2.80in]{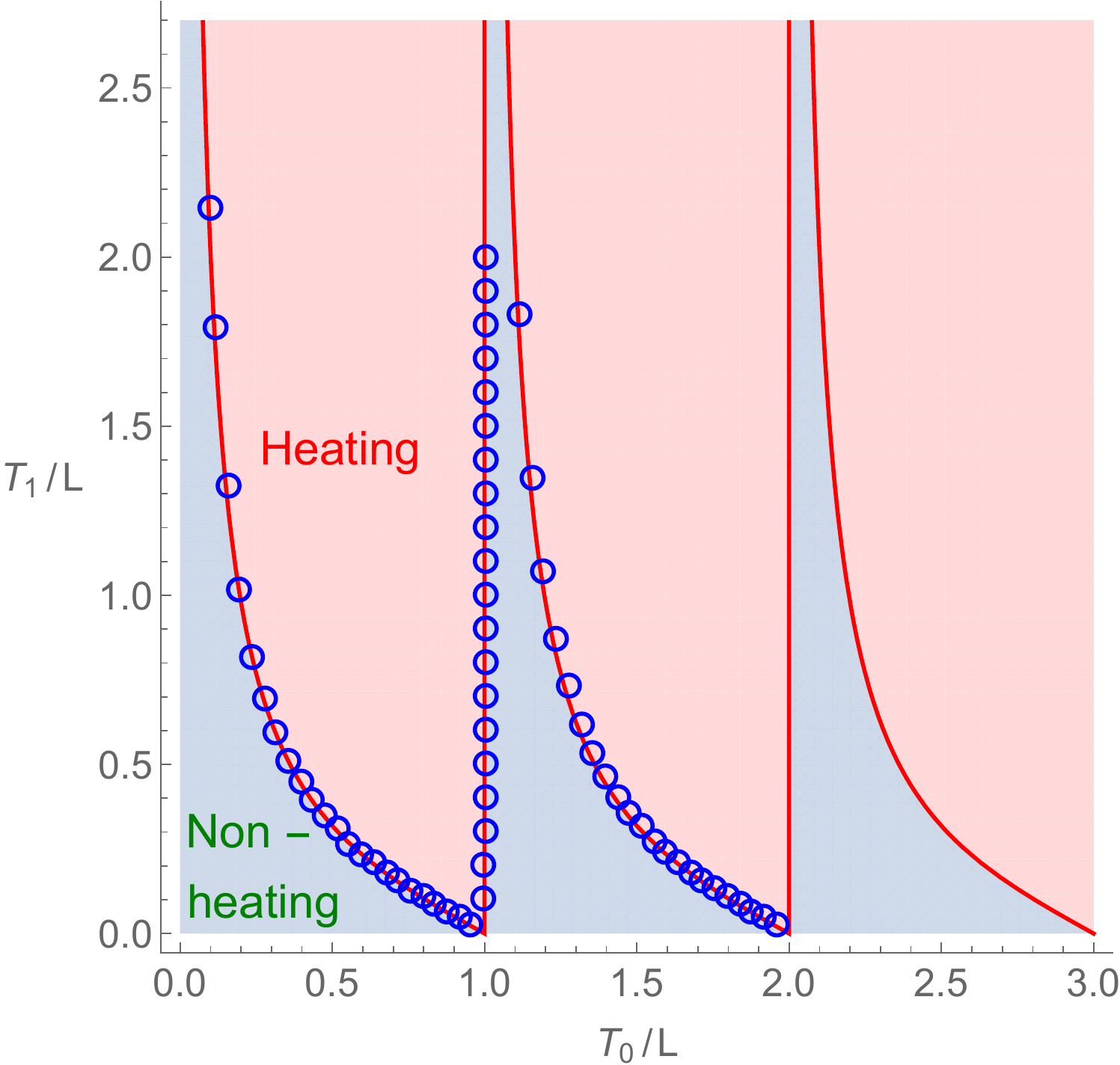}
\caption{(Part of) Phase diagram for a Floquet CFT, plotted according to Eq.\eqref{ThreeMobiusTransformation}.
The solid dots are obtained from a numerical simulation based on a free fermion chain with $L=500$ and $T_0/L<2$.
}
\label{PhaseDiagram}
\end{figure}

\textbf{\textit{One-cycle driving}} $\,\,$
Before studying the effect of $n$-cycle driving, it is helpful to
check how a primary operator $\mathcal{O}$ evolves under one-cycle driving.
First, with a conformal mapping
$z=e^{\frac{2\pi w}{L}}$,
we map the strip in $w$-plane to the complex $z$-plane, where the boundaries along $x=0$ and $x=L$ in $w$-plane
are mapped to the slit along the half real axis $\text{Re}(z)\ge 0$ in $z$-plane.
Based on the study in Ref.\onlinecite{WenWu2018}, one can find that under one-cycle driving,
the operator $\mathcal{O}$ in $z$-plane evolves from $(z,\bar{z})$ to $(z_1,\bar{z}_1)$ as follows
\be\label{U_eff}
U_{\text{eff}}^{\dag}\,\mathcal{O}(z,\bar{z})\,U_{\text{eff}}=
\left(
\frac{\partial z_1}{\partial z}
\right)^h\left(\frac{\partial \bar{z}_1}{\partial \bar{z}}\right)^{\bar{h}}\mathcal{O}(z_1,\bar{z}_1),
\ee
where we have defined the time evolution operator $U_{\text{eff}}:=e^{-H_0\tau_0}e^{-H_1\tau_1}$,
\footnote{It is noted that studying $U_{\text{eff}}$
is equivalent to studying the property of the Floquet Hamiltonian $H_F$, which is defined through
$U_{\text{eff}}=e^{-H_F(\tau_0+\tau_1)}$. Aside from the types of M$\ddot{\text{o}}$bius transformations in Eq.\eqref{z1},
one can alternatively use the Floquet Hamiltonian to characterize/classify different phases.
A detailed discussion on the Floquet Hamiltonian and its spectrum in a Floquet CFT will be given in \onlinecite{WenUN}.
}
and $h$ ($\bar{h}$) is the conformal dimension of $\mathcal{O}$.
In particular, $z_1$ ($\bar{z}_1$) is related to $z$ ($\bar{z}$) by a M$\ddot{\text{o}}$bius
transformation: \cite{WenWu2018,MobiusWiki}
\be\label{z1}
z_{1}=f(z)=\frac{az+b}{cz+d} \quad \rightarrow \quad \mathfrak{H}=
\left(
\begin{split}
& a &b\\
&c  &d
\end{split}
\right),
\ee
where we have associated to the M$\ddot{\text{o}}$bius transformation $z_1=f(z)$ a matrix $\mathfrak{H}$,
and defined $a:=(1+\frac{\pi \tau_1}{L})\cdot e^{\frac{\pi \tau_0}{L}}$,
$b:=-\frac{\pi \tau_1}{L}\cdot e^{\frac{-\pi \tau_0}{L}}$,
$c:=\frac{\pi \tau_1}{L}\cdot e^{\frac{\pi \tau_0}{L}}$,
and $d:=\left(1-\frac{\pi \tau_1}{L}\right)\cdot e^{\frac{-\pi \tau_0}{L}}$.
As will be seen shortly, \textit{the M$\ddot{\text{o}}$bius transformation in Eq.\eqref{z1} determines
the Floquet dynamics of our periodically driven CFT}.
Note that the M$\ddot{\text{o}}$bius transformation has been normalized so that $ad-bc=1$.
By defining the trace square of $\mathfrak{H}$ as
$\sigma(\mathfrak{H}):=\big[\text{Tr}(\mathfrak{H})\big]^2$, it is known
that the value of $\sigma(\mathfrak{H})$ classifies different types M$\ddot{\text{o}}$bius transformations.
After analytical continuation $\tau_0\to iT_0$ and $\tau_1\to iT_1$, one can find that
\be\label{TraceSquare}
\sigma(\mathfrak{H})=4(1-\Delta),
\ee
with
\be\label{Delta}
\Delta=\Big[1-\left(\frac{\pi T_1}{L}\right)^2\Big]\sin^2\frac{\pi T_0}{L}+
\frac{\pi T_1}{L}\cdot \sin\frac{2\pi T_0}{L}.
\ee
Depending on the values of $\Delta$, there are three types of
M$\ddot{\text{o}}$bius transformations as follows:\cite{MobiusWiki}
\be\label{ThreeMobiusTransformation}
\left\{
\begin{split}
&0\le \sigma<4, \quad &0<\Delta\le 1,\quad &\text{Elliptic},\\
&\sigma=4, \quad &\Delta=0, \quad&\text{Parabolic},\\
&\sigma>4, \quad &\Delta<0, \quad &\text{Hyperbolic}.
\end{split}
\right.
\ee
The elliptic, parabolic, and hyperbolic transformations
are conjugate to the operation of rotation, translation, and dilation in $z$-plane, respectively.
As we will see, $\Delta=0$ determines the phase transition in a Floquet CFT (see Fig.\ref{PhaseDiagram}),
and the elliptic and hyperbolic transformations correspond to non-heating and heating phases in the phase diagram,
respectively.

\begin{figure}[tp]
\includegraphics[width=3.00in]{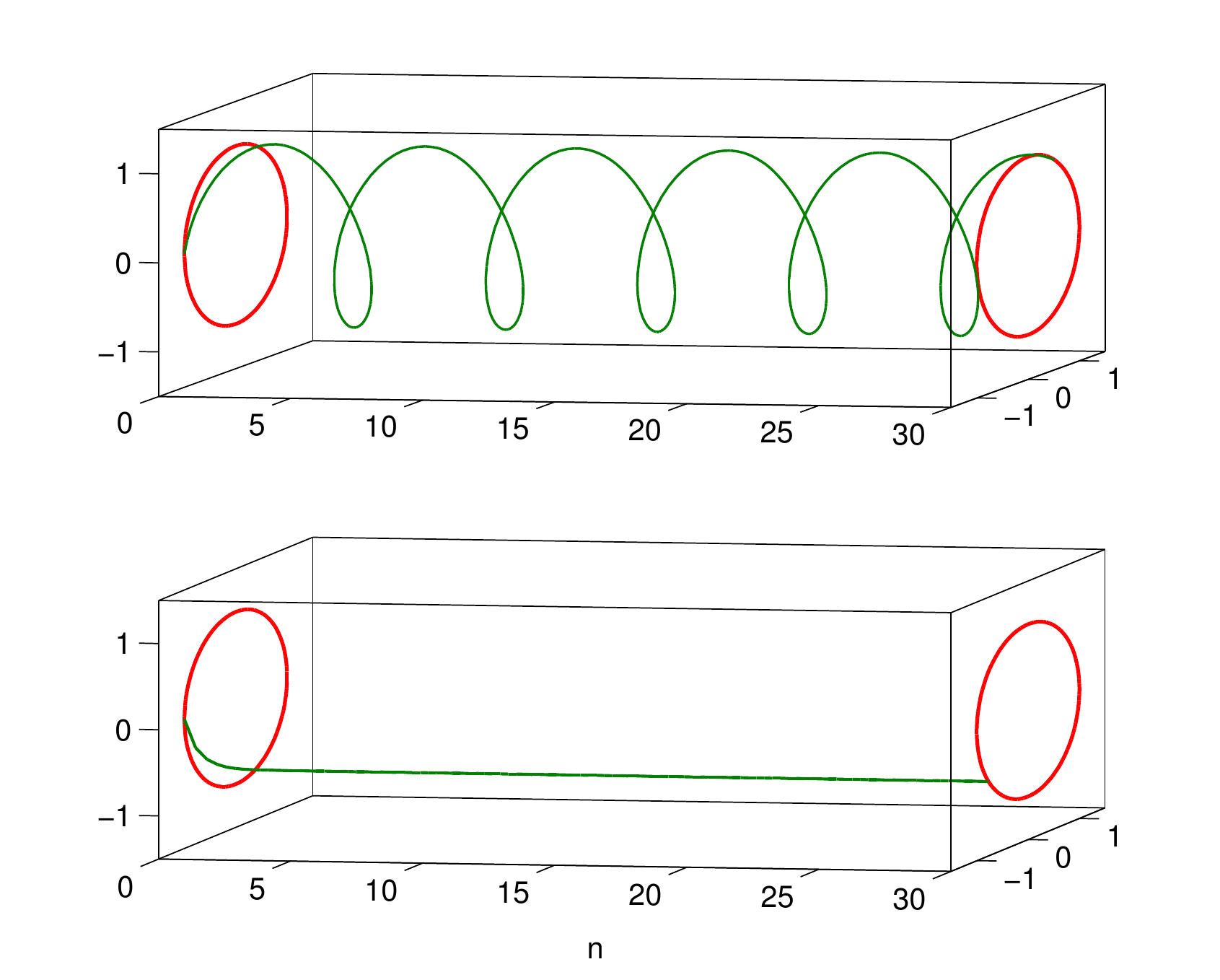}
\caption{Trajectory of $z_n$ in $z$-plane as a function of $n$ in the non-heating phase (top) and the heating phase (bottom).
We choose $T_0=T_1=L/10$ in the non-heating phase and $T_0=T_1=L/2$ in the heating phase.
The subsystem length is chosen as $l=L/2$.}
\label{TrajectoryBoth}
\end{figure}

\textbf{\textit{$n$-cycle driving}}$\,\,$
To have a more intuitive picture on the effect of different M$\ddot{\text{o}}$bius
transformations in Eq.\eqref{ThreeMobiusTransformation},
let us consider the $n$-cycle driving.
It can be found that the operator $\mathcal{O}$ in $z$-plane, after $n$ cycles of driving,
is driven from $(z,\bar{z})$ to $(z_n,\bar{z}_n)$, with
\be\label{NormalForm}
\frac{z_n-\gamma_1}{z_n-\gamma_2}=\eta^n \cdot\frac{z-\gamma_1}{z-\gamma_2},
\ee
and similarly for $\bar{z}_n$.
Here $\gamma_1$ and $\gamma_2$ are called `fixed points' of the M$\ddot{\text{o}}$bius transformation
and are determined by $f(\gamma)=\gamma$ in
Eq.\eqref{z1}, with the explicit expression $\gamma_{1,2}=\big(a-d\mp \sqrt{(a-d)^2+4bc}\big)/2c$. The multiplier
$\eta$ shows qualitatively different behaviors depending on the types of M$\ddot{\text{o}}$bius
transformations (after analytical continuation):
\be
\eta=\left\{
\begin{split}
&e^{i2\phi},\quad &\text{Elliptic},\\
&1, \quad &\text{Parabolic},\\
&e^{2\phi'},\quad &\text{Hyperbolic}.
\end{split}
\right.
\ee
$\phi$ and $\phi'$ are real functions of driving periods $T_0$ and $T_1$, and the explicit expressions will be
given in the following discussions on entanglement entropy.
Several remarks here: First, as shown in Fig.\ref{TrajectoryBoth},
for $\Delta >0$, \textit{i.e.}, the M$\ddot{\text{o}}$bius transformation
is elliptic, $\eta$ is a phase, and one can find that the trajectory of $z_n$ in the complex $z$-plane keeps
oscillating as a function of $n$.
\footnote{It is noted that although the trajectory is plotted in a continuous way, it is
only well defined at discrete $n$. It is the same in the following
plots for entanglement entropy evolution $S_A(t)$, where $t$ is defined at discrete values $t=n(T_0+T_1)$.
}
On the other hand, for $\Delta<0$, \textit{i.e.}, the M$\ddot{\text{o}}$bius transformation is hyperbolic, $z_n$ will
converge to one of the fixed points $\gamma_{1,2}$ (depending on $\phi'>0$ or $\phi'<0$)
 exponentially in $n$, and will not come back to its initial value.
This difference will result in different behaviors of correlation functions and entanglement
entropy evolution. Second, for $\Delta=0$, \textit{i.e.}, the M$\ddot{\text{o}}$bius transformation is parabolic,
one can find that $\eta=1$ and the two fixed points merge into a single one, namely $\gamma_1=\gamma_2=\gamma=(a-d)/2c$.
In this case, one cannot use Eq.\eqref{NormalForm} to determine the trajectory of $z_n$. It can be found that
$z_n$ is now determined by
\be\label{Zn_parabolic}
\frac{1}{z_n-\gamma}=\frac{1}{z-\gamma}+n\cdot \beta,
\ee
where $\beta=c$ [see the expression of $c$ below Eq.\eqref{z1}] is the so-called `translation length'.
Then, $z_n$ converges to the fixed point $\gamma$ in the way $z_n-\gamma\propto \frac{1}{n}$ for large $n$,
in contrast to the exponential convergence in the hyperbolic case.
As a remark, it is interesting to compare our (1+1)-d Floquet CFT with the (0+1)-d quantum Mathieu's
harmonic oscillator. These two systems have similar phase diagrams due to the underlying
 algebra structures which are isomorphic to each other.
\footnote{
In our setup for the (1+1)-d Floquet CFT, the trajectory of $\mathcal{O}(z_n,\bar{z}_n)$ in $z$-plane
displays three kinds of behaviors depending on the types of M$\ddot{\text{o}}$bius transformations.
This is similar to certain classical Floquet dynamics such as the classical Mathieu's harmonic oscillator
(see, \textit{e.g.}, Ref.\onlinecite{kawai2002parametrically}), where
the harmonic oscillator displays three kinds of
trajectories in the phase space depending on
the elliptic, parabolic or hyperbolic transformations between $(x_n, p_n)^T$ and
$(x_{n+1}, p_{n+1})^T$. $x_n$ and $p_n$ are the position and
momentum of the harmonic oscillator after $n$ cycles of driving.
Depending on the trajectories of the harmonic oscillator, there are stable and non-stable regions separated by
a boundary, similar to the non-heating and heating phases with a phase transition in our Floquet CFTs.
(For the non-stable region in Mathieu's harmonic oscillator, the amplitude of oscillator keeps increasing by absorbing
energy from the external driving. This is similar to the heating phase of our Floquet CFTs, where
the entanglement entropy keeps growing in time.)
In addition, the `phase diagram' of Mathieu's oscillator also shows periodic structure as the driving period increases,
which results from higher order resonances.
In fact, there is a deep reason on the similarity between our (1+1)-d Floquet CFTs and the
(0+1)-d quantum Mathieu's harmonic oscillators.
The Hamiltonians in our Floquet CFTs are composed of three generators of $sl(2,R)$ algebra, while
the Hamiltonians in \textit{quantum} Mathieu's harmonic oscillators are composed of three generators
of $su(1,1)$ algebra.\cite{perelomov1969group}
The similarity on the `phase diagram' of the two systems originates from
the algebraic structure $sl(2,R)\cong su(1,1)$, where `$\cong$' represents `is isomorphic to'.
From this point of view, we may say that within our setup a (1+1)-d Floquet CFT $\cong$ a (0+1)-d
quantum Mathieu's harmonic oscillator.
}

\begin{figure}[tp]
\includegraphics[width=2.8in]{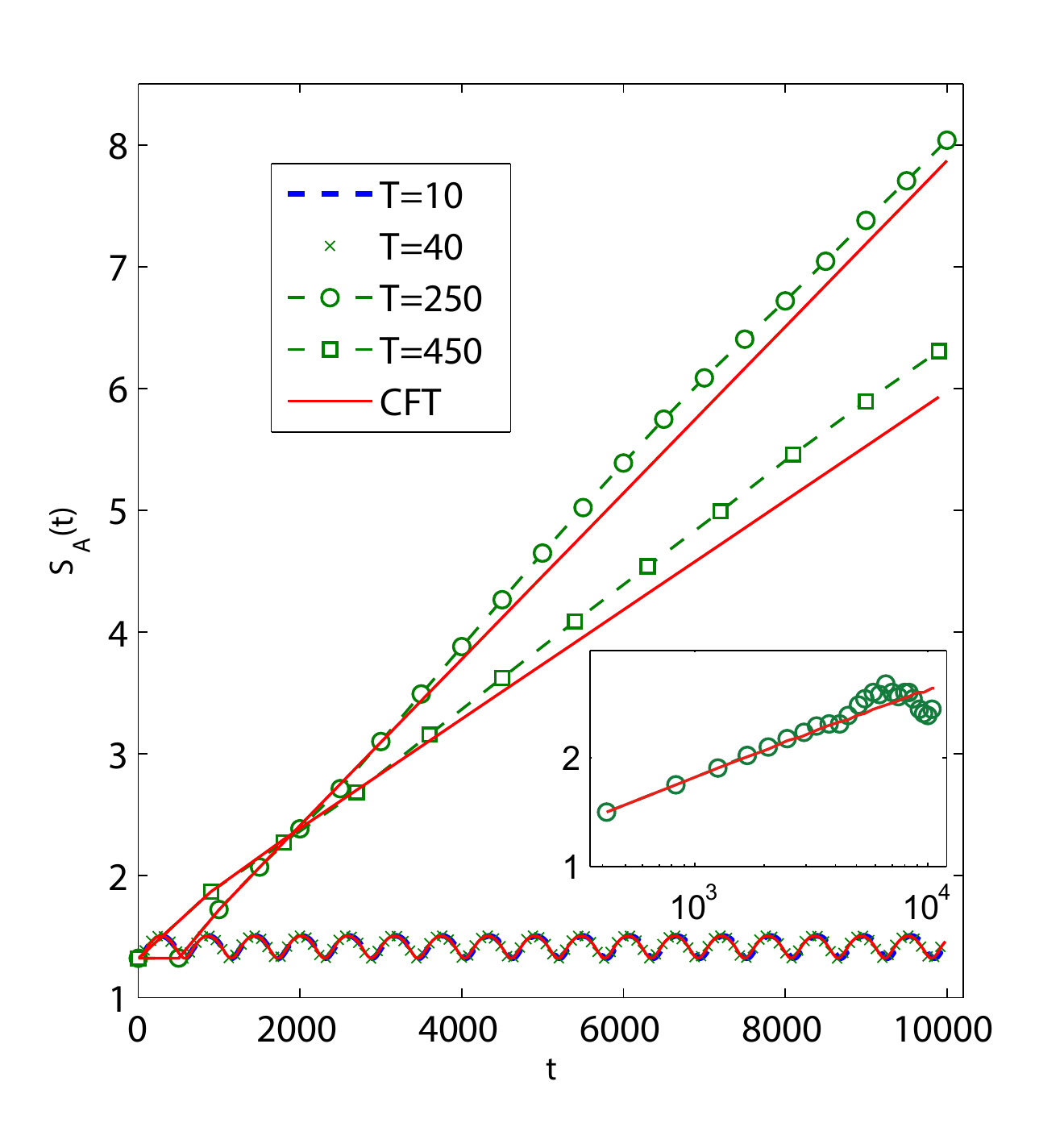}
\caption{
Comparison of entanglement entropy evolution
between CFT calculations and numerical simulations in heating phase,
non-heating phase, and at the phase transition (inset). We choose $T_0=T_1=T$, $L=500$ and $l=L/2$.
The phase transition happens at $T^{\ast}/L\simeq 0.416$ in CFT prediction and at
$T^{\ast}/L\simeq 0.418$ in the numerical simulation for $T<L$.
}
\label{CompareCFT2}
\end{figure}

\begin{table*}
\begin{ruledtabular}
\begin{tabular}{ccccccc}
          &Mobius Transformation  &Conjugate to  &Multipliers   &   Entanglement growth  &Single-point function & Phases in Floquet CFT\\ \hline
 & Elliptic     & Rotation & $\eta=e^{i2\phi}$ & Oscillating  & Oscillating & Non-heating   \\ \hline
&Parabolic  &Translation  &$\eta=1$ & Logarithmic & Power-law decay & Phase transition \\ \hline
&Hyperbolic & Dilation  &$\eta=e^{2\phi'}$ & Linear &Exponential decay & Heating \\
\end{tabular}
\end{ruledtabular}
\caption{Summary of correspondence between M$\ddot{\text{o}}$bius transformations and different phases (and phase transition) in a Floquet CFT.}
\end{table*}

\textbf{\textit{Single-point function and Entanglement entropy}}$\,\,$
To characterize the Floquet dynamics, now let us focus on
two physical quantities, \textit{i.e.}, single-point correlation functions and entanglement entropy for a subsystem $A=[0,l]$.
For a primary operator $\mathcal{O}$, one has
$\langle \psi(t)|\mathcal{O}(w,\bar{w})|\psi(t)\rangle
=
\left(\frac{\partial z}{\partial w}\right)^h
\left(\frac{\partial \bar{z}}{\partial \bar{w}}\right)^{\bar{h}}
\left(\frac{\partial z_n}{\partial z}\right)^h
\left(\frac{\partial \bar{z}_n}{\partial \bar{z}}\right)^{\bar{h}}
\langle \mathcal{O}(z_n,\bar{z}_n)\rangle_z$,
where $\langle\cdots\rangle_z$ represents the correlation function of ``$\cdots$'' in $z$-plane, and
$ \langle \mathcal{O}(z_n,\bar{z}_n)\rangle_z
=A^{\text{b}}_{\mathcal{O}}\cdot (\frac{1}{4}
\frac{1}{\sqrt{z_n}}\frac{1}{\sqrt{\bar{z}_n}})^h \cdot
(\frac{2\epsilon i}{\sqrt{z_n}-\sqrt{\bar{z}_n}})^{2h}$.
Here $A_{\mathcal{O}}^{\text{b}}$ is an amplitude depending on the selected boundary condition $|b\rangle$,
and $\epsilon$ is a UV cutoff which may be interpreted as the lattice constant in a lattice model.
Then the $\alpha$-th Renyi entanglement entropy $S^{(\alpha)}_A(t)$ for $A=[0,l]$ is directly
related to the single point correlation function of
twist operator $\mathcal{T}_{\alpha}$, which is itself a primary operator with conformal dimension
$h=\bar{h}=\frac{c}{24}\left(\alpha-\frac{1}{\alpha}\right)$, where $c$ is the central charge and $\alpha$
is the Renyi index. Explicitly, one has\cite{calabrese2004entanglement,calabrese2009entanglement}
\be\label{SA}
S^{(\alpha)}_A(t)=\frac{1}{1-\alpha}\log\langle \psi(t)|\mathcal{T}_{\alpha}|\psi(t)\rangle,
\ee
where $\mathcal{T}_{\alpha}$ is inserted at $w=0+il$  in the $w$-plane.
In later discussion, we will use the von Neumann entropy defined by
$S_A(t):=\lim_{\alpha\to 1}S^{(\alpha)}_A(t)$.
Based on Eq.\eqref{SA}, one can infer the behavior of single-point correlation function from $S^{(\alpha)}_A(t)$ (and
vice versa),\cite{SM}
and therefore we will mainly focus on the entanglement entropy hereafter.
In \onlinecite{SM}, we have obtained the analytical expression of $S_A(t)$ for $A=[0,l]$ with
$l\in (0,L)$ under arbitrary driving periods $T_0$ and $T_1$. Since the expression is quite involved in general,
as an illustration, we will mainly focus on $l=L/2$, for which the entanglement entropy has an elegant expression.

\textbf{\textit{Non-heating phase}}$\,\,$
In the non-heating phase, both the correlation functions and the entanglement entropy keep oscillating in time.
This phase corresponds to the case $\Delta>0$ in Eq.\eqref{TraceSquare} and the
M$\ddot{\text{o}}$bius transformation is elliptic.
One can find the entanglement entropy of subsystem $A=[0,L/2]$ as
\be\label{SA_nonheating}
\boxed{
S_A(t)\simeq\frac{c}{6}\log\Big[\frac{L}{\pi\epsilon}\cdot
\frac{R\cos(2n\phi+\varphi)-K}{R\cos\varphi-K}
\Big]}\,,\quad \text{for}\,\, \Delta>0,
\ee
where the driving time is defined as $t:=n(T_0+T_1)$.
Here (and in the following) we use ``$\simeq$'' instead of ``$=$'' because we only keep the leading term of $S_A(t)$.
The subleading constant term that depends on the boundary condition is of order $\mathcal{O}(1)$ and is neglected hereafter.
The parameters in Eq.\eqref{SA_nonheating} depend on the driving periods $T_0$ and $T_1$ as follows:
$e^{i2\phi}:=(Q-iP)(Q+iP)$, with $Q:=\sin\frac{\pi T_0}{L}-\frac{L}{\pi T_1}\cos\frac{\pi T_0}{L}$,
and $P:=\frac{L}{\pi T_1}\sqrt{\Delta}$.
$Re^{i\varphi}:=W\cos\frac{\pi T_0}{L}+1+iP\sin\frac{\pi T_0}{L}$ and $K:=W\cos\frac{\pi T_0}{L}+W^2$,
with $W:=\cos\frac{\pi T_0}{L}+\frac{L}{\pi T_1}\sin\frac{\pi T_0}{L}$.
For $n=0$, \textit{i.e.}, there is no driving, one has $S_A(t=0)=\frac{c}{6}\log\frac{\pi L}{\epsilon}$, which is the entanglement entropy in the ground state of $H_0$, as expected.

\begin{figure}[tp]
\includegraphics[width=2.9in]{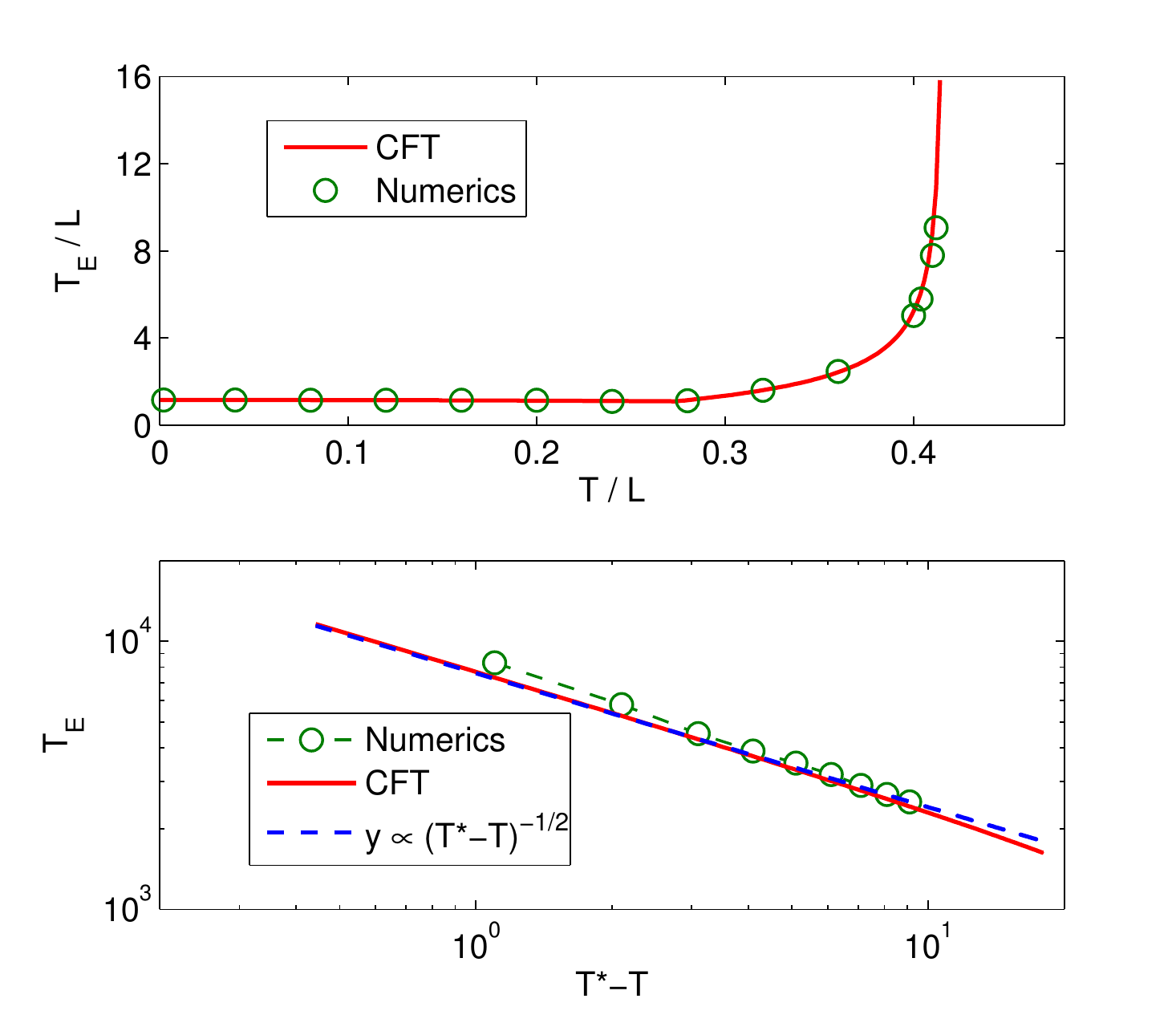}
\caption{
(Top) Oscillation period $T_E$ of the entanglement entropy for $A=[0,L/2]$
as a function of the driving period $T$. (We choose
$T_0=T_1=T$.) (Bottom) Scaling behavior of $T_E$ near the phase transition.
}
\label{OscillationCriticalPoint}
\end{figure}

There are several remarkable features for $S_A(t)$ in Eq.\eqref{SA_nonheating}:
(i) $S_A(t)$ oscillates as a function of driving cycles $n$ all the way
(and so does the single-point correlation function\cite{SM}), indicating that the system is not heated.
The oscillation period of entanglement entropy in time $t$ is
\be\label{T_E}
T_E=\pi\cdot(T_0+T_1)/|\phi|.
\ee
(ii) In the high-frequency driving limit $T_0,T_1\ll L$, $S_A(t)$ only depends on the ratio of $T_1$ and $T_0$.\cite{SM}
 Here let us take $T_0=T_1=T$, then one can find that in the
limit $T\ll L$, $S_A(t)$ has a simple form
\be\label{SA_oscillation00b}
S_A(t)\simeq \frac{c}{6}\log \frac{L}{\pi \epsilon}+\frac{c}{6}\log\big[
2-\cos(\sqrt{3}\pi t/L)\big],
\ee
where $t:=n(T_0+T_1)=2nT$. Then the oscillation period of $S_A(t)$ is $T_E=2L/\sqrt{3}$,
which is proportional to the length of the system.
In other words, in the high frequency limit, $T_E$ is independent of the driving frequency, as shown in Fig.\ref{OscillationCriticalPoint}.
To further understand the result in Eq.\eqref{SA_oscillation00b},
we note that in the high frequency limit $T\ll L$, one may consider
the approximation $e^{-H_0T}e^{-H_1T}\simeq e^{-(H_0+H_1)T}$.
Then the high-frequency driving limit of Floquet dynamics corresponds to a single quench with the effective Hamiltonian
$H_F=\frac{1}{2}(H_0+H_1)$, which has been studied in
in Ref.\onlinecite{WenWu2018}. Therein it was found that the entanglement entropy evolution indeed
displays oscillations with period $T_E=2L/\sqrt{3}$.\cite{WenWu2018} (See also \onlinecite{SM} for more discussions.)

\begin{figure}[tp]
\includegraphics[width=2.8in]{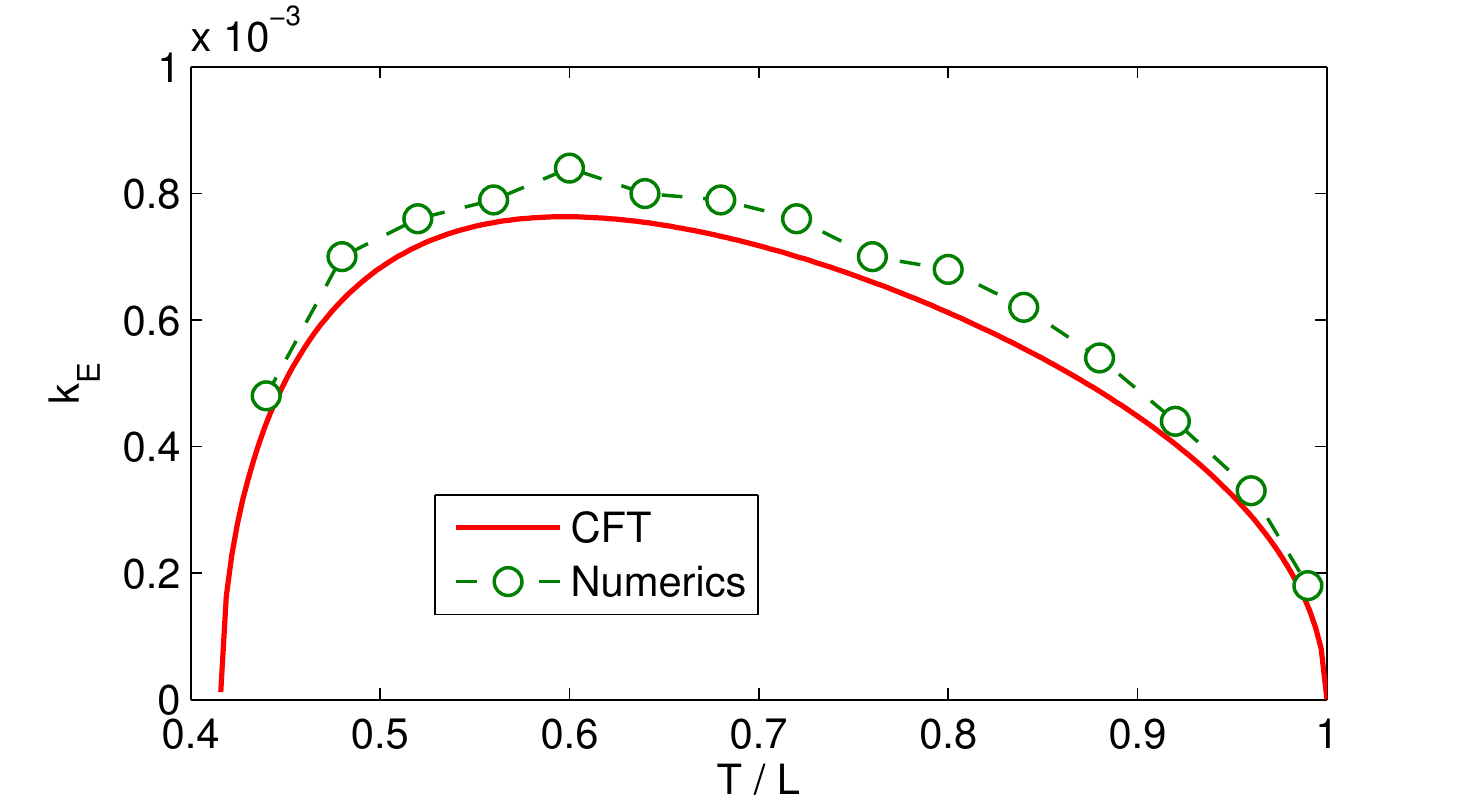}
\caption{The slope $k_E$ of linear growth in $S_A(t)$ as a function of $T$ in the heating phase.
We choose $T_0=T_1=T$ and $L=500$ in the numerical simulations.
The phase transitions happen at  $T^{\ast}/L\simeq 0.416$ and $T^{\ast}/L=1$ for $T\le 1$.
}
\label{SlopeChange}
\end{figure}

\textbf{\textit{Heating phase}}$\,\,$
In contrast to the non-heating phase, the entanglement entropy keeps growing in time in the heating phase,
and the single-point function decays exponentially in time.
This phase corresponds to the case $\Delta<0$ in Eq.\eqref{TraceSquare} and the Mobius transformation is
hyperbolic. The entanglement entropy for $A=[0,L/2]$ has the expression
\be\label{SA_heating}\boxed{
S_A(t)\simeq \frac{c}{6}\log\Big[
\frac{L}{\pi \epsilon}\cdot\frac{R'\cosh(2n\phi'+\varphi')-K}{R'\cosh \varphi'-K}\Big]}\,,
\, \text{for}\,\,\Delta<0.
\ee
As before, here we neglect the subleading constant term.
The parameters in Eq.\eqref{SA_heating} are as follows.
$e^{2\phi'}:=(Q+P)/(Q-P)$, where $Q$ has the same expression as the non-heating case, \textit{i.e.},
$Q:=\sin\frac{\pi T_0}{L}-\frac{L}{\pi T_1}\cos\frac{\pi T_0}{L}$, but $P$ is now expressed as
$P:=\frac{L}{\pi T_1}\sqrt{-\Delta}$.
$R'e^{\varphi'}:=W\cos\frac{\pi T_0}{L}+1-P\sin\frac{\pi T_0}{L}$,
$R'e^{-\varphi'}:=W\cos\frac{\pi T_0}{L}+1+P\sin\frac{\pi T_0}{L}$,
and
$K:=W\cos\frac{\pi T_0}{L}+W^2$, with $W:=\cos\frac{\pi T_0}{L}+\frac{L}{\pi T_1}\sin\frac{\pi T_0}{L}$.
Compared to $S_A(t)$ for the non-heating phase in Eq.\eqref{SA_nonheating},
all the parameters are defined similarly except that $\Delta\to -\Delta$ and therefore $P\to iP$.
The `$\cos$' term in Eq.\eqref{SA_nonheating} now
becomes a `$\cosh$' term, which may be intuitively viewed as a transition from `real time' to
`imaginary time'.
Again, for $n=0$, $S_A(t)$ reduces to the ground state entropy.
As $n$ grows, $S_A(t)$ can be approximated as
\be\label{LinearGrowth}
S_A(t)\simeq \frac{c}{6}\log \frac{L}{\pi \epsilon}+\frac{c}{3}\cdot \frac{|\phi'|}{T_0+T_1}\cdot t, \quad
t=n(T_0+T_1).
\ee
I.e., the entanglement entropy grows linearly in time $t$ [see Fig.\ref{CompareCFT2} for a typical plot
based on Eq.\eqref{SA_heating}], with slope
\be
k_E:=\frac{\partial S_A(t)}{\partial t}=\frac{c}{3}\cdot \frac{|\phi'|}{T_0+T_1}.
\ee
We emphasize that here $S_A(t)$ keeps growing all the way since there are
infinite number of degrees of freedom and the energy spectrum goes to infinity with no upper bound in CFT.
\footnote{It is noted that the UV cutoff $\epsilon$ in $S_A(t)$ is introduced only at the entanglement cut.
In the bulk of the subsystem $A$, there are always infinite degrees of freedom in a quantum field theory.
And the energy spectrum (in a CFT) goes to infinity without an upper bound. Then the system can
keep absorbing energy.
Similar things also appear in the entanglement entropy in a CFT with finite temperature.
In the high temperature limit, the entanglement entropy for a finite subsystem of length $l$ is
$S_A(\beta)\simeq \frac{c}{3}\log \frac{l}{\epsilon}+\frac{c}{3}\cdot \frac{\pi l}{\beta}$,
 where $\epsilon$ is the UV cutoff introduced at the entanglement cut.\cite{cardy2016entanglement}
The entanglement entropy grows linearly with the temperature $1/\beta$ all the way.
This is not the case in a lattice, where there are always a finite number of degrees of freedom in a finite subsystem,
and the bandwidth of energy spectrum is finite.
The entanglement entropy in a lattice system will finally saturate as $1/\beta$ increases.
}
In a lattice model, however, the entanglement entropy will finally
saturate because of the UV cutoff introduced by the lattice constant.\cite{SM}
As shown in Fig.\ref{SlopeChange}, we plot the slope of the linear growth, \textit{i.e.}, $k_E$,
as a function of $T$ (by choosing $T_0=T_1=T$).
It can be found that the slope goes to zero as we approach the phase transitions,
which indicates that the linear-growth behavior disappears at the phase transitions, as expected.

\textbf{\textit{Phase Transition}}
The phase transition between heating and non-heating phases
happens at $\Delta=0$,
where the entanglement entropy grows logarithmically in time, and the single-point function decays in a power-law
in time. There are two sets of solutions for $\Delta=0$ [see Eq.\eqref{Delta}].
One is
$
T_0=mL$, with $ m=1, 2, 3,\cdots,$
as depicted in the vertical lines in Fig.\ref{PhaseDiagram}.
The entanglement entropy for $A= [0,\, L/2]$ has a simple expression
\be\label{SAlogMiddle01}
S_A(t)\simeq \frac{c}{6}\log\Big\{
\frac{L}{\pi \epsilon}\cdot
\Big[
1+4n^2\cdot\left(\frac{\pi T_1}{L}\right)^2
\Big]\Big\},
\ee
where $t:=n(T_0+T_1)=n(mL+T_1)$. In the large $n$ limit, one has $S_A(t)\simeq \frac{c}{3}\log t$.
Another set of solutions for $\Delta=0$ are determined by
$\Big[1-\left(\frac{\pi T_1}{L}\right)^2\Big]\sin\frac{\pi T_0}{L}+
2\cdot \frac{\pi T_1}{L}\cdot \cos\frac{\pi T_0}{L}=0$.
In this case, one has
\be\label{SAlogSolution2}
S_A(t)\simeq \frac{c}{6}\log \Bigg\{\frac{L}{\pi \epsilon}\Big(
\frac{4 \left(\frac{\pi T_1}{L}\right)^4}{1+\left(\frac{\pi T_1}{L}\right)^2}\, n^2
-\frac{4\left(\frac{\pi T_1}{L}\right)^2}{1+\left(\frac{\pi T_1}{L}\right)^2}\, n+1
\Big)\Bigg\}.\nonumber
\ee
Different from $S_A(t)$ in Eq.\eqref{SAlogMiddle01}, now $S_A(t)$ decreases first and then grows in time.
Again, in the large $n$ limit, one has $S_A(t)\simeq \frac{c}{3}\log t$.
A typical plot for the entanglement entropy at the phase transitions can be found in Fig.\ref{PhaseTransitionIandII}.

The correspondence between different phases and M$\ddot{\text{o}}$bius transformations \textit{et al.}
is summarized in Table I.

\textbf{\textit{Near the phase transition}}
Now let us check the entanglement entropy evolution near the phase transition.
First, as we approach the phase transition from the non-heating phase, as shown in Fig.\ref{OscillationCriticalPoint},
one can find that the oscillation period of $S_A(t)$ diverges.
By taking $T_1=k\cdot T_0$ with arbitrary $k>0$,
one can find that\cite{SM}
\be
T_E\propto |T_0-T_0^{\ast}|^{-1/2}.
\ee
The critical exponent is independent of $k$.
If we approach the phase transition from the heating phase, as shown in Fig.\ref{SlopeChange},
the slope $k_E$ of the linear growth will vanish. In other words, $1/k_E$ will diverge, and we find that\cite{SM}
\be
1/k_E\propto |T_0-T_0^{\ast}|^{-1/2}.
\ee
In short, by approaching the phase transition from both sides, one can obtain the
critical exponent $\zeta=1/2$.

\textbf{\textit{Comparison with numerics}} $\,\,$
We compare our CFT calculation with the numerical simulations based on a free fermion lattice
which has finite sites $L$ with open boundary conditions.
We prepare the initial state as the ground state of
$H_0=\frac{1}{2}\sum_{i=1}^{L-1}c_i^{\dag}c_{i+1}+h.c.$ with half filling.
The sine-square deformed Hamiltonian has the form
$H_1=\sum_{i=1}^{L-1}\sin^2\left(\frac{\pi (i+1/2)}{L}\right)
c_i^{\dag}c_{i+1}+h.c.$,
where $c_i$ ($c_i^{\dag}$) are fermionic operators, which satisfy the
anticommutation relations $\{c_i,c_j\}=\{c_i^{\dag},c_j^{\dag}\}=0$, and
$\{c_i,c_j^{\dag}\}=\delta_{ij}$.
We compare our field theory result with the numerical simulations in
Figs.\ref{PhaseDiagram}, \ref{CompareCFT2}, \ref{OscillationCriticalPoint} and \ref{SlopeChange}, respectively.
The agreement in the non-heating phase is remarkable.
In the heating phase, the numerical results deviate from the CFT results as $t$ grows.
This is as expected, since the lattice system can no longer be well described by a CFT as it keeps absorbing energy.
(Recall that only the low energy limit can be well described by a CFT.)

\textbf{\textit{Discussion and Conclusion}}$\,\,$
We have proposed an analytically solvable setup to study the Floquet dynamics of a generic CFT.
The phase diagram, entanglement entropy and correlation functions can be analytically obtained.
There are many future problems, and we mention a few of them:
(i) The Hamiltonians $H_0$ and $H_1$ considered in this work are composed of three generators of
$sl(2,R)$ algebra, which is a subalgebra of the Virasoro algebra in a two dimensional CFT.
Our setup applies to the general case with $H(t)=H(t+T)$,
as long as $H(t)$ is a combination of the three generators of $sl(2,R)$ algebra, \textit{i.e.},
the Virasoro generators $L_0$, $L_n$ and $L_{-n}$ (and the anti-holomorphic parts),
as will be discussed in more detail in Ref.\onlinecite{WenUN}.
On the other hand, it is an open question if the Hamiltonian $H(t)$ is a combination of generators of the Virasoro algebra,
which is infinite dimensional.
(ii) It is also desirable to study the multi-point correlation functions (although quite involved) in our setup.
As discussed in \onlinecite{SM}, our system with periodic driving is not uniformly heated. It is our
future work to use two-point correlation functions to measure the local `temperature' of the Floquet CFT.
(iii) Our setup also works for non-periodic driving schemes, such as the
quasi-periodic driving and random driving CFTs, which deserve future studies.
(iv) Since our setup applies to a generic CFT including the large-$c$ CFT,
it would also be interesting to consider the holographic description of our setup\cite{RT2006,RT2006b,HRT}, 
which may shed new lights on the Floquet dynamics in AdS/CFT.\cite{auzzi2013periodically,rangamani2015driven,Mas1712}

\textbf{\textit{Acknowledgement}}$\,\,$
XW thanks Shinsei Ryu and  Andreas W. W. Ludwig for introducing to him
the concept of SSD of a CFT in the collaboration in \onlinecite{Ryu1604}.
We thank for helpful conversations and discussions with Zhen Bi, Po-Yao Chang,
Yingfei Gu, Max Metlitski,
Xiao-Liang Qi, Yang Qi, Cecile Repellin, Shinsei Ryu, and Xiao-Gang Wen,
and thank Liujun Zou for many helpful comments on various aspects of our results.
We also thank for the helpful comments and questions during the seminar talk at MIT.
XW is supported by the Gordon and Betty Moore Foundation's EPiQS initiative
through Grant No. GBMF4303 at MIT. JQW is supported by Massachusetts Institute
of Technology and  the Simons foundation it from qubit collaboration.

\let\oldaddcontentsline\addcontentsline
\renewcommand{\addcontentsline}[3]{}
\bibliography{Floquet}
\let\addcontentsline\oldaddcontentsline

\begin{center}
\textbf{\large
Floquet CFT:\\
 Supplemental Materials
}
\end{center}

\tableofcontents

\section{On the setup and entanglement entropy}

In the supplementary materials, we present more details on the calculations,
as well as analysis and discussions on the results in the main text.

\subsection{More about the setup}

The system is driven by two different Hamiltonians periodically, with
\be
\left\{
\begin{split}
H_0=&\int_0^L h(x) dx,\\
H_1=&\int_0^L\sin^2\left(\frac{\pi x}{L}\right) h(x)dx,
\end{split}
\right.
\ee
where $h(x)$ is the Hamiltonian density which is uniform in space. We start from the ground state of $H_0$, and drive the
system with $H_1$ and $H_0$ periodically (see the main text).
In the CFT calculation, it is convenient to
rewrite the Hamiltonian in terms of stress energy tensors, \textit{i.e.},
\be
\left\{
\begin{split}
H_0=&\int_0^L \frac{dx}{2\pi}T_{\tau\tau}(x),\\
H_1=&H_0-\frac{1}{2}\left(H_++H_-\right),
\end{split}
\right.
\ee
with $T_{\tau\tau}=T(w)+\bar{T}(\bar{w})$, and
\be
H_{\pm}=\int_0^L\frac{dx}{2\pi}
\left(
e^{\pm 2\pi w/L}T(w)+e^{\mp2\pi\bar{w}/L}\bar{T}(\bar{w})
\right).
\ee
The CFT lives on a finite space of length $L$, with conformal boundary condition imposed.
In path integral, the partition function is defined on a strip
\be
w=\tau+ix,
\ee
where $\tau$ is the imaginary time, and $x$ is the space, with
\be
\tau\in(-\infty, +\infty), \quad x\in (0,L).
\ee
The wavefunction after $n$ cycles of driving can be written as
\be
|\psi(t)\rangle=e^{-iH_0T_0}e^{-iH_1T_1}\cdots
e^{-iH_0T_0}e^{-iH_1T_1}
|G\rangle.
\ee
We can evaluate the $n$-point (equal time) correlation function in state $|\psi(t)\rangle$ as follows:
\be
\langle \psi(t)|\mathcal{O}_1 \mathcal{O}_2\cdots \mathcal{O}_n|\psi(t)\rangle.
\ee
To obtain the correlation functions of operators at different time, we simply need to insert these operators
at different time slices.
Shown in Fig.\ref{FloquetPI} is the path integral representation of single-point correlation function
$\langle \psi(t)|\mathcal{O}(x)|\psi(t)\rangle$.
In the calculation, we will consider the Euclidean space, \textit{i.e.},
\be
|\psi(\tau)\rangle=e^{-H_0\tau_0}e^{-H_1\tau_1}\cdots
e^{-H_0\tau_0}e^{-H_1\tau_1}
|G\rangle,
\ee
and take analytical continuation $\tau_0\to iT_0$ and $\tau_1\to iT_1$ in the final step.
We map the strip $w$-plane to $z$-plane as follows,
\begin{equation}
\begin{gathered}
\includegraphics[width=8cm]{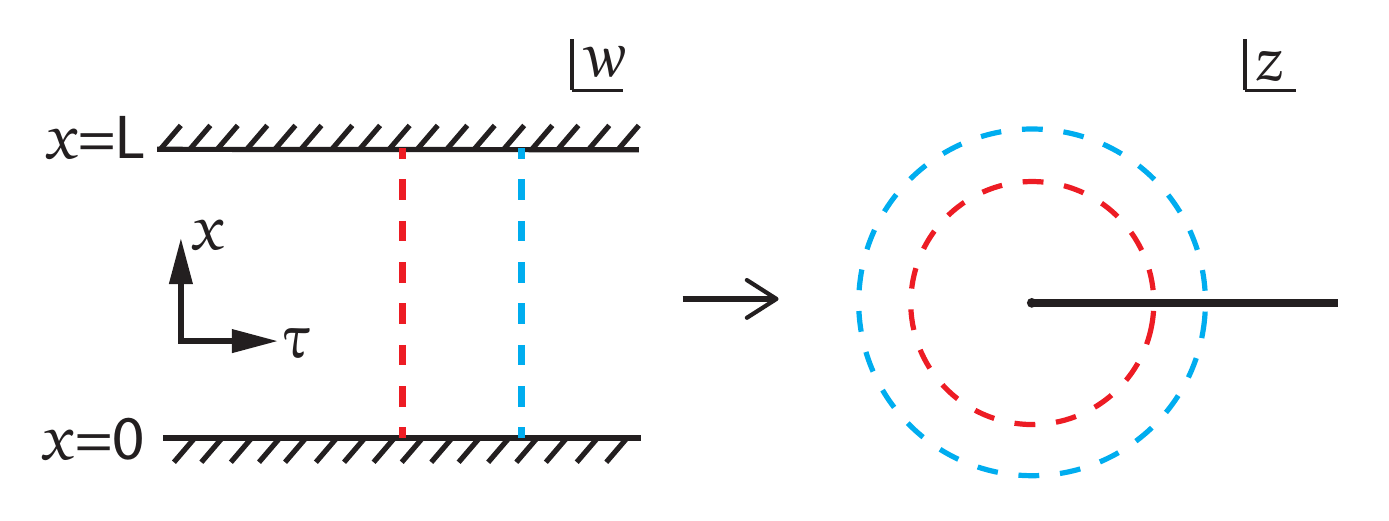}
\end{gathered}\nonumber
\end{equation}
by using the conformal transformation
\be
z=e^{\frac{2\pi w}{L}}.
\ee
Instead of studying how $e^{-H_0\tau_0}$ and $e^{-H_1\tau_1}$ act on the ground state, we
consider the Heisenberg picture here. I.e., we study how the operator evolves
during the periodic driving.
For the operator $\mathcal{O}(z,\bar{z})$ in $z$-plane, it is found that the effect of Hamiltonian
$H_i$, with $i=0,1$, is to evolve the operator in the following way
\be
e^{\tau H_i}\mathcal{O}(z,\bar{z})e^{\tau H_i}=\left(\frac{\partial z^i_{\text{new}}}{\partial z}\right)^h
\left(\frac{\partial \bar{z}^i_{\text{new}}}{\partial \bar{z}}\right)^h\mathcal{O}(z^i_{\text{new}},\bar{z}^i_{\text{new}}).
\ee
To be concrete, we have\cite{WenWu2018}
\be
\left\{
\begin{split}
z^0_{\text{new}}=&e^{\frac{2\pi \tau_0}{L}}z,\\
z^1_{\text{new}}=&\frac{(1+\frac{\pi \tau_1}{L})z-\frac{\pi \tau_1}{L}}{\frac{\pi \tau_1}{L}z+\left(1-\frac{\pi \tau_1}{L}\right)},
\end{split}
\right.
\ee
Then, after one-cycle driving, one can obtain
\be
U_{\text{eff}}^{\dag}\,\mathcal{O}(z,\bar{z})\,U_{\text{eff}}=
\left(
\frac{\partial z_1}{\partial z}
\right)^h\left(\frac{\partial \bar{z}_1}{\partial \bar{z}}\right)^{\bar{h}}\mathcal{O}(z_1,\bar{z}_1),
\ee
where we have defined the time evolution operator $U_{\text{eff}}:=e^{-H_0\tau_0}e^{-H_1\tau_1}$.
$z_1$ has the explicit expression
\be
z_1=
\frac{(1+\frac{\pi \tau_1}{L})\cdot e^{\frac{2\pi \tau_0}{L}}\cdot z-\frac{\pi \tau_1}{L}}{\frac{\pi \tau_1}{L}\cdot e^{\frac{2\pi \tau_0}{L}}\cdot z+\left(1-\frac{\pi \tau_1}{L}\right)},
\ee
Written in a normalized form of M$\ddot{\text{o}}$bius transformation, $z_1$ has the expression:
\be\label{z1_SM}
z_1=
\frac{(1+\frac{\pi \tau_1}{L})\cdot e^{\frac{\pi \tau_0}{L}}\cdot z-\frac{\pi \tau_1}{L}\cdot e^{\frac{-\pi \tau_0}{L}}
}{\frac{\pi \tau_1}{L}\cdot e^{\frac{\pi \tau_0}{L}}\cdot z+\left(1-\frac{\pi \tau_1}{L}\right)\cdot e^{\frac{-\pi \tau_0}{L}}}
=:\frac{az+b}{cz+d}.
\ee
That is, we have defined $a$, $b$, $c$ and $d$ as follows:
\be\label{abcd_SM}
\left\{
\begin{split}
a:=&(1+\frac{\pi \tau_1}{L})\cdot e^{\frac{\pi \tau_0}{L}},\\
b:=&-\frac{\pi \tau_1}{L}\cdot e^{\frac{-\pi \tau_0}{L}},\\
c:=&\frac{\pi \tau_1}{L}\cdot e^{\frac{\pi \tau_0}{L}},\\
d:=&\left(1-\frac{\pi \tau_1}{L}\right)\cdot e^{\frac{-\pi \tau_0}{L}},
\end{split}
\right.
\ee
which satisfies $ad-bc=1$.
Note that $\tau_0$ and $\tau_1$ are real numbers, and $\bar{z}_1$ has the form
\be
\bar{z}_{1}=
\frac{(1+\frac{\pi \tau_1}{L})\cdot e^{\frac{\pi \tau_0}{L}}\cdot \bar{z}-\frac{\pi \tau_1}{L}\cdot e^{\frac{-\pi \tau_0}{L}}
}{\frac{\pi \tau_1}{L}\cdot e^{\frac{\pi \tau_0}{L}}\cdot \bar{z}+\left(1-\frac{\pi \tau_1}{L}\right)\cdot e^{\frac{-\pi \tau_0}{L}}}.
\ee
This M$\ddot{\text{o}}$bius transformation (before analytical continuation) forms a $SL(2,R)$ group.
For later convenience, we write the M$\ddot{\text{o}}$bius transformation in Eq.\eqref{z1_SM} in the normal form:
\be\label{Normal_1cycle_SM}
\frac{z_1-\gamma_1}{z_2-\gamma_2}=\eta\cdot\frac{z-\gamma_1}{z-\gamma_2},
\ee
where $\gamma_1$ and $\gamma_2$ are called `fixed points', and $\eta$ is called `multiplier' in
a M$\ddot{\text{o}}$bius transformation.
$\gamma_1$, $\gamma_2$ and $\eta$ have the explicit form
\be
\left\{
\begin{split}
\gamma_{1}=&\frac{a-d- \sqrt{(a-d)^2+4bc}}{2c},\\
\gamma_{2}=&\frac{a-d+ \sqrt{(a-d)^2+4bc}}{2c},\\
\eta=&\frac{
(a+d)+\sqrt{(a-d)^2+4bc}
}{
(a+d)-\sqrt{(a-d)^2+4bc}
}.
\end{split}
\right.
\ee
Note that one can take an inverse on both sides of Eq.\eqref{Normal_1cycle_SM}, so that
$\gamma_1\to \gamma_2$, $\gamma_2\to \gamma_1$ and $\eta\to \eta^{-1}$.

Now we repeat the above procedure for $n$ cycles of driving,
and denote the coordinate of $\mathcal{O}$ as $z_n$ and $\bar{z}_n$. Then one has
\be
\langle \psi(\tau)|\mathcal{O}(z,\bar{z})|\psi(\tau)\rangle=
\left(\frac{\partial z_n}{\partial z}\right)^h
\left(\frac{\partial \bar{z}_n}{\partial \bar{z}}\right)^{\bar{h}}\mathcal{O}(z_n,\bar{z}_n),
\ee
where
\be\label{z_n_SM}
\left\{
\begin{split}
\frac{z_n-\gamma_1}{z_n-\gamma_2}=&\eta^n\cdot \frac{z-\gamma_1}{z-\gamma_2},\\
\frac{\bar{z}_n-\gamma_1}{\bar{z}_n-\gamma_2}=&\eta^n\cdot \frac{\bar{z}-\gamma_1}{\bar{z}-\gamma_2}.
\end{split}
\right.
\ee

\subsection{Expression for entanglement entropy}
\label{EE_expression}

The entanglement entropy of subsystem $A=[0,l]$ with $0<l<L$ can be obtained
by calculating the single-point correlation function of a twist operator $\mathcal{T_{\alpha}}$.
The entanglement measure we use is the so-called Renyi entropy
\be\label{RenyiEE}
S_A^{(\alpha)}(t)=\frac{1}{1-\alpha}\log \text{tr}\left[\rho_A^{\alpha}(t)\right],
\ee
where $\alpha$ is the Renyi index,
and the von Neumann entropy
\be
S_A(t)=\lim_{\alpha\to 1}S_A^{(\alpha)}(t).
\ee
The term $\text{tr}(\rho_A^{\alpha})$ in $S_A^{(\alpha)}(t)$ is related with the single-point correlation function
of twist operator as follows:\cite{calabrese2004entanglement,calabrese2009entanglement}
\be\label{Tn}
\text{tr}(\rho_A^{\alpha})=\langle\psi(t)| \mathcal{T}_{\alpha}(x=l) |\psi(t)\rangle,
\ee
where $\mathcal{T}_{\alpha}$ is a primary operator with conformal dimension
\be\label{ConformalDimension}
h=\bar{h}=\frac{c}{24}\left(\alpha-\frac{1}{\alpha}\right),
\ee
In the following, we will evaluate the correlation function in Eq.\eqref{Tn} with path integral method.
Pictorially, $\langle \psi(t)|\mathcal{T}_{\alpha}(x=l)|\psi(t)\rangle$ is shown in Fig.\ref{FloquetPI}
by setting $O(x)=\mathcal{T}_{\alpha}(x)$.
Note that there are both Euclidean time and Lorentzian time in the path integral.
As shown in the following, we will do calculation in the Euclidean space by setting $it=\tau$, and
analytically continue back to Lorentzian time in the final step.

Let us start by evaluating the single-point correlation function:
\be\label{Oww_SM}
\begin{split}
&\langle \psi(t)|\mathcal{T}_{\alpha}(w,\bar{w})|\psi(t)\rangle\\
=&
\left(\frac{\partial z}{\partial w}\right)^h
\left(\frac{\partial \bar{z}}{\partial \bar{w}}\right)^{h}
\left(\frac{\partial z_n}{\partial z}\right)^h
\left(\frac{\partial \bar{z}_n}{\partial \bar{z}}\right)^{h}
\langle \mathcal{T}_{\alpha}(z_n,\bar{z}_n)\rangle_z,
\end{split}
\ee
where we have considered the fact $h=\bar{h}$ for the twist operator.
Note that $\langle \mathcal{T}_{\alpha}(z_n,\bar{z}_n)\rangle_z$
is the single point correlation function in the ground state in $z$-plane, with a slit lying
along the half real-axis $[0,\infty)$. Conformal boundary conditions are imposed along the slit.
Then one has
\be\label{OnePoint}
 \langle \mathcal{T}_{\alpha}(z_n,\bar{z}_n)\rangle_z
=A_{\alpha}^{\text{b}}\cdot (\frac{1}{4}z_n^{-\frac{1}{2}}\bar{z}_n^{-\frac{1}{2}})^h \cdot
(\frac{2\epsilon i}{z_n^{\frac{1}{2}}-\bar{z}_n^{\frac{1}{2}}})^{2h},
\ee
where $A^{\text{b}}_{\alpha}$ is an amplitude depending on the selected boundary condition $|b\rangle$
as well as the Renyi index $\alpha$.
It will affect the entanglement entropy by an order $\sim\mathcal{O}(1)$ term.
$\epsilon$ is a UV cut-off, which may be considered as the lattice spacing in a lattice model.
In Eq.\eqref{Oww_SM}, one has
\be
\left(\frac{\partial z}{\partial w}\right)^h
\left(\frac{\partial \bar{z}}{\partial \bar{w}}\right)^{h}=\left(\frac{2\pi}{L}\right)^{2h}.
\ee
To evaluate other terms in Eq.\eqref{Oww_SM},
first we rewrite Eq.\eqref{z_n_SM} as
\be
z_n=\frac{
(\gamma_1-\eta^n \gamma_2)z-(1-\eta^n)\gamma_1\gamma_2
}{
(1-\eta^n)z-(\gamma_2-\eta^n \gamma_1)
}
=:\frac{\mathfrak{a}z+\mathfrak{b}}{\mathfrak{c}z+\mathfrak{d}},
\ee
where we have defined
\be\label{abcd_Ncycle}
\left\{
\begin{split}
\mathfrak{a}:=&\gamma_1-\eta^n \gamma_2,\\
\mathfrak{b}:=&-(1-\eta^n)\gamma_1\gamma_2,\\
\mathfrak{c}:=&1-\eta^n,\\
\mathfrak{d}:=&-(\gamma_2-\eta^n \gamma_1).
\end{split}
\right.
\ee
One can find that
\be
\left(\frac{\partial z_n}{\partial z}\right)^h\left(\frac{\partial \bar{z}_n}{\partial \bar{z}}\right)^h
=\left[
\frac{(\mathfrak{a}\mathfrak{d}-\mathfrak{b}\mathfrak{c})^2}{(\mathfrak{c}^2+2\mathfrak{c}\mathfrak{d}\cdot \cos\frac{2\pi l}{L}+\mathfrak{d}^2)^2}
\right]^h
\ee
We also need to evaluate the single-point correlation function
$ \langle \mathcal{T}_{\alpha}^{(z)}(z_n,\bar{z}_n)\rangle$
 in Eq.\eqref{OnePoint}.
For convenience, we write $z_n$ as
\be
z_n=\frac{E+iF}{G}=:R_n \cdot e^{i\varphi_n},
\ee
where
\be
\left\{
\begin{split}
E=&\mathfrak{a}\mathfrak{c}+(\mathfrak{a}\mathfrak{d}+\mathfrak{b}\mathfrak{c})\cos\frac{2\pi l}{L}+\mathfrak{b}\mathfrak{d},\\
F=&(\mathfrak{a}\mathfrak{d}-\mathfrak{b}\mathfrak{c})\sin\frac{2\pi l}{L},\\
G=&\mathfrak{c}^2+2\mathfrak{c}\mathfrak{d}\cdot \cos\frac{2\pi l}{L}+\mathfrak{d}^2.
\end{split}
\right.
\ee
Then, we have
\be\label{OnePoint002}
\begin{split}
& \langle \mathcal{T}_{\alpha}^{(z)}(z_n,\bar{z}_n)\rangle\\
=&A_{\alpha}^{\text{b}}\cdot (\frac{1}{4}z_n^{-\frac{1}{2}}\bar{z}_n^{-\frac{1}{2}})^h \cdot
(\frac{2\epsilon i}{z_n^{\frac{1}{2}}-\bar{z}_n^{\frac{1}{2}}})^{2h},\\
=&A_{\alpha}^{\text{b}}\cdot \left(\frac{1}{4}\cdot \frac{1}{\sqrt{z_n\bar{z}_n}}\right)^h \cdot
\left(\frac{-4 \epsilon^2 }{z_n+\bar{z}_n-2\sqrt{z_n\bar{z}_n}}\right)^{h},\\
=&A_{\alpha}^b\cdot \left(\frac{1}{4}\cdot \frac{1}{R_n}\right)^h\cdot
\left(
\frac{-4\epsilon^2}{2R_n\cdot \cos\varphi_n-2R_n}
\right)^h\\
=&A_{\alpha}^b\cdot \left[
\frac{\epsilon^2}{2}\cdot \frac{1}{R_n^2\cdot (1-\cos\varphi_n)}
\right]^h.
\end{split}
\ee
Note that
\be
\left\{
\begin{split}
R_n:=&\Big(\frac{E^2+F^2}{G^2}\Big)^{1/2},\\
\cos\varphi_n:=&\frac{E/G}{\sqrt{(E^2+F^2)/G^2}}.
\end{split}
\right.
\ee
Collecting all these terms, one can obtain [see Eq.\eqref{Oww_SM}]
\be\label{Oww002}
\begin{split}
&\langle \psi(t)|\mathcal{T}_{\alpha}(w,\bar{w})|\psi(t)\rangle\\
=&
\left(\frac{\partial z}{\partial w}\right)^h
\left(\frac{\partial \bar{z}}{\partial \bar{w}}\right)^h
\left(\frac{\partial z_n}{\partial z}\right)^h
\left(\frac{\partial \bar{z}_n}{\partial \bar{z}}\right)^h
\langle \mathcal{O}^{(z)}(z_n,\bar{z}_n)\rangle\\
=&
\left(\frac{2\pi}{L}\right)^{2h}\cdot
\left[\frac{(\mathfrak{a}\mathfrak{d}-\mathfrak{b}\mathfrak{c})^2}{(\mathfrak{c}^2
+2\mathfrak{c}\mathfrak{d}\cdot \cos\frac{2\pi l}{L}+\mathfrak{d}^2)^2}\right]^h\\
&\cdot
A_{\alpha}^b\cdot \left[
\frac{\epsilon^2}{2}\cdot \frac{1}{R_n^2\cdot (1-\cos\varphi_n)}
\right]^h\\
=&
A_{\alpha}^b\cdot \left(\frac{2\pi}{L}\right)^{2h}\cdot
\left[\frac{\epsilon^2}{2}\cdot
\frac{(\mathfrak{a}\mathfrak{d}-\mathfrak{b}\mathfrak{c})^2}{E^2+F^2-\frac{\sqrt{G^2}}{G}E\sqrt{E^2+F^2}}\right]^h.
\end{split}
\ee
Then, based on Eqs.\eqref{RenyiEE} and \eqref{Tn}, the $\alpha$-th Renyi entropy is related with $\langle\psi(t)|\mathcal{T}_{\alpha}(w,\bar{w})|\psi(t)\rangle$ as follows
\be
S_A^{(\alpha)}=\frac{1}{1-\alpha}\log\langle\psi(t)|\mathcal{T}_{\alpha}(w,\bar{w})|\psi(t)\rangle.
\ee
One can find the entanglement entropy as
\be\label{SA002}
S_A(t)\simeq
\frac{c}{6}\log \frac{L}{\pi \epsilon}+\frac{c}{12}\log \frac{E^2+F^2
-\frac{\sqrt{G^2}}{G}E\sqrt{E^2+F^2}}{2(\mathfrak{a}\mathfrak{d}-\mathfrak{b}\mathfrak{c})^2},
\ee
where we only keep the leading term, and the subleading term of order $\mathcal{O}(1)$ has been neglected.
In the following parts, we need to evaluate Eqs.\eqref{Oww002} and \eqref{SA002},
by making analytical continuation $\tau_0\to iT_0$ and $\tau_1\to iT_1$.

\section{Entanglement entropy evolution}

As discussed in the main text, the behavior of the entanglement entropy
evolution is determined by the sign of $\Delta$ in Eq.\eqref{Delta}.
For $\Delta>0$, we have the non-heating phase. Both the entanglement entropy
and the single-point correlation function oscillate in time;
For $\Delta<0$, we have the heating phase. The entanglement entropy keeps growing
linearly in time, and the single-point correlation function decays exponentially in time;
For $\Delta=0$, there is a phase transition. The entanglement entropy grows
logarithmically in time, and the single point correlation function shows a power-law
decay in time.

In the following, we give the explicit expressions of $S_A(t)$ in Eq.\eqref{SA002}
for different cases,  by doing analytical continuation $\tau_0\to iT_0$ and $\tau_1\to iT_1$.
The procedure is tedious but quite straightforward, and here we list the
main results and give some discussions.

\subsection{Non-heating phase}

\begin{figure}[tp]
\includegraphics[width=3.0in]{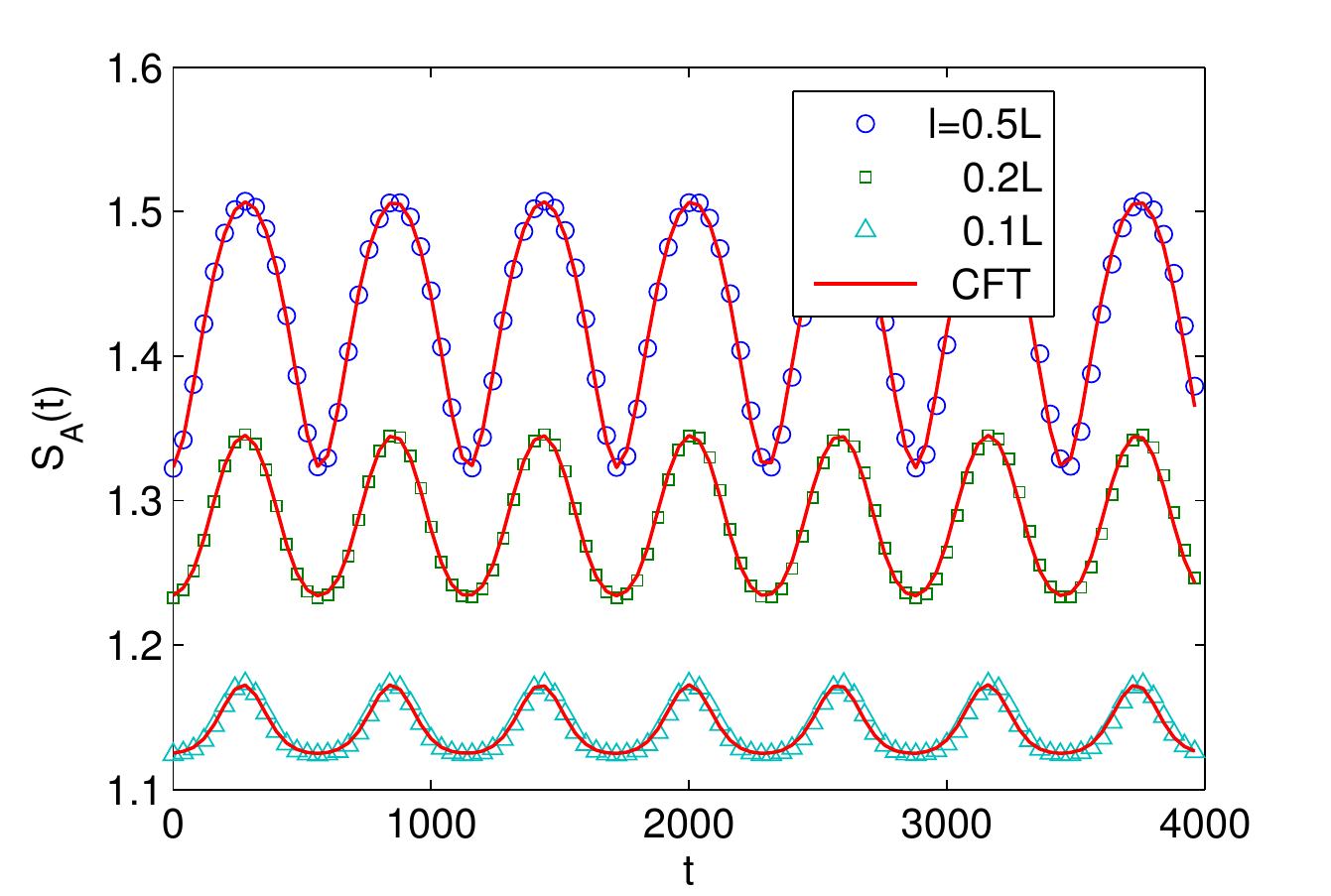}
\caption{
Comparison between numerical simulations and CFT calculations for
entanglement evolution in the non-heating phase with $c=1$.
Here we choose $L=500$, and $T_0=T_1=T=20$.
}
\label{T20_DifferentLA}
\end{figure}

The non-heating phase corresponds to $\Delta>0$ in Eq.\eqref{Delta}.
After doing analytical continuation, one can find
\be\label{SA_nonheating_SM}
S_A(t)\simeq\frac{c}{6}\log\frac{L}{\pi \epsilon}
+\frac{c}{12}\,\log\frac{\mathcal{E}^2+\mathcal{F}^2-\mathcal{E}\sqrt{\mathcal{E}^2+\mathcal{F}^2}}{2P^4},
\ee
with $t:=n(T_0+T_1)$ and
\be\label{EF_nonheating0A_SM}
\left\{
\begin{split}
\mathcal{E}=&R\cos(2n\phi+\varphi)-K,\\
\mathcal{F}=&P^2\sin\frac{2\pi l}{L},
\end{split}
\right.
\ee
where we have defined
\be\label{Parameters_SM}
\left\{
\begin{split}
e^{i2\phi}:=&(Q-iP)(Q+iP),\\
Q:=&\sin\frac{\pi T_0}{L}-\frac{L}{\pi T_1}\cos\frac{\pi T_0}{L},\\
P:=&\frac{L}{\pi T_1}\sqrt{\Delta},\\
K:=&W\cos\frac{\pi T_0}{L}-W^2\cos\frac{2\pi l}{L},\\
W:=&\cos\frac{\pi T_0}{L}+\frac{L}{\pi T_1}\sin\frac{\pi T_0}{L},\\
Re^{i\varphi}:=&W\cos\frac{\pi T_0}{L}-\cos\frac{2\pi l}{L}+iP\sin\frac{\pi T_0}{L}.\\
\end{split}
\right.
\ee
One can find that in the non-heating phase, the entanglement entropy oscillates in time, with the period
\be\label{TE_osc_SM}
T_E=\frac{\pi}{|\phi|}\cdot(T_0+T_1).
\ee
Now let us do a self-consistent check. For $n=0$, \textit{i.e.}, the system is not driven at all and stays in the ground
state, one can find that
\be
\begin{split}
\mathcal{E}=&R\cos\varphi-K
=(W^2-1)\cos\frac{2\pi l}{L}\\
=&P^2\cos\frac{2\pi l}{L},
\end{split}
\ee
where we have considered the fact that $W^2-P^2=1$.
Then one can obtain
\be
S_A(t=0)=\frac{c}{6}\log\left(\frac{L}{\pi\epsilon}\sin\frac{\pi l}{L}\right),
\ee
which is the entanglement entropy in the ground state, as expected.

For a generic $l$, a typical plot deep in the non-heating phase is shown in Fig.\ref{T20_DifferentLA}.
It can be found that $S_A(t)$ for different $l$ oscillate with the same period, as can be also
straightforwardly observed in Eqs. \eqref{SA_nonheating_SM}$\sim$\eqref{Parameters_SM}.

Now let us check the specific case $l=L/2$ as discussed in the main text.
In this case, one has
$\mathcal{E}=R\cos(2n\phi+\varphi)-K$, and $\mathcal{F}=0$.
Moreover, one can find that
\be
|R|-K<0,\quad \mathcal{E}<0,
\ee
based on the following facts:
$R^2=\left(W+\cos\frac{\pi T_0}{L}\right)^2$,
$K^2=W^2\cdot \left(W+\cos\frac{\pi T_0}{L}\right)^2$,
$W^2=P^2+1>1$, and
$K>0$.
Note that $K>0$ because $K=W^2+W\cos\frac{\pi T_0}{L}>W^2-W^2=0$.

Therefore, the entanglement entropy in Eq.\eqref{SA_nonheating_SM} becomes
\be
S_A(n)=\frac{c}{6}\log\frac{L}{\pi\epsilon}+\frac{c}{6}\log\frac{
K-R\cos(2n\phi+\varphi)
}{P^2}.
\ee
Note also that $K-R\cos\varphi=P^2$.
Therefore, one has
\be
S_A(n)=\frac{c}{6}\log \Big[\frac{L}{\pi\epsilon}\cdot\frac{
K-R\cos(2n\phi+\varphi)
}{K-R\cos\varphi}\Big],
\ee
which is Eq.\eqref{SA_nonheating} in the main text.

\subsubsection{High frequency limit}

Now let us look at the behavior of $S_A(t)$ in the high frequency limit $T_0,\,T_1\ll L$.
We will show that the result only depends on the ratio
\be
\sigma:=T_0/T_1.
\ee
In this limit, one can find that the parameters in Eq.\eqref{Parameters_SM}
can be approximated by (keeping the leading order)
$\Delta\simeq \left(\frac{\pi T_0}{L}\right)^2+\frac{\pi T_1}{L}\left(\frac{2\pi T_0}{L}\right)$,
$P\simeq \sqrt{\left(\frac{T_0}{T_1}\right)^2+\frac{2T_0}{T_1}}=\sqrt{\sigma^2+2\sigma}$,
$Q\simeq -\frac{L}{\pi T_1}$,
$W\simeq 1+\frac{T_0}{T_1}$,
$K\simeq\left(1+\frac{T_0}{T_1}\right)-\left(1+\frac{T_0}{T_1}\right)^2\cos\frac{2\pi l}{L}$,
$\phi\simeq\frac{\pi T_1}{L}\cdot \sqrt{\left(\frac{T_0}{T_1}\right)^2+\frac{2T_0}{T_1}}$, and
$Re^{i\varphi}\simeq 1+\frac{T_0}{T_1}-\cos\frac{2\pi l}{L}$.
Then $\mathcal{E}$, $\mathcal{F}$ and $P$ in Eq.\eqref{SA_nonheating_SM} are approximated by
\be\label{EFP_general}
\left\{
\begin{split}
\mathcal{E}\simeq& \left(1+\sigma-\cos\frac{2\pi l}{L}\right)\cdot \cos\left(2n\cdot \frac{\pi T_1}{L}\cdot \sqrt{\sigma^2+2\sigma}\right)\\
&-(1+\sigma)+(1+\sigma)^2\cos\frac{2\pi l}{L},\\
\mathcal{F}\simeq& (\sigma^2+2\sigma)\cdot\sin\frac{2\pi l}{L},\\
P\simeq&\sqrt{\sigma^2+2\sigma}.
\end{split}
\right.
\ee
From Eqs.\eqref{SA_nonheating_SM} and \eqref{EFP_general}, one can obtain
the oscillation period of entanglement entropy as
\be
T_E=\frac{1+\sigma}{\sqrt{\sigma^2+2\sigma}}L.
\ee
In particular, for $l=L/2$, the entanglement entropy has a simple expression
\be
S_A(t)\simeq \frac{c}{6}\log\frac{L}{\pi \epsilon}+\frac{c}{6}
\log\frac{1+\sigma-\cos\left(\frac{\sqrt{\sigma^2+2\sigma}}{1+\sigma}\cdot \frac{2\pi t}{L}\right)}{\sigma}.
\ee
Now let us consider the simple case $T_0=T_1=T$. Then one has
\be\label{EFP}
\left\{
\begin{split}
\mathcal{E}\simeq &\left(2-\cos\frac{2\pi l}{L}\right)\cdot \cos\left(\frac{\sqrt{3}\pi}{L}t\right)-2+4\cos\frac{2\pi l}{L},\\
\mathcal{F}\simeq &3\sin\frac{2\pi l}{L},\\
P\simeq& \sqrt{3},
\end{split}
\right.
\ee
where $t=2T$.
As a self-consistent check, one can find that for $t=0$, one has
$\mathcal{E}\simeq 3\cos\frac{2\pi l}{L}$.
Then $S_A(t)$ in Eq.\eqref{SA_nonheating_SM} has the expression
$S_A(t=0)=\frac{c}{6}\log\left(\frac{L}{\pi \epsilon}\cdot \sin\frac{\pi l}{L}\right)$,
which is the entanglement entropy in the ground state, as expected.
From Eqs.\eqref{SA_nonheating_SM} and \eqref{EFP}, one can find that
the oscillation period of entanglement entropy is
\be
T_E=\frac{2L}{\sqrt{3}},
\ee
which is observed in the numerical simulation in Fig.\ref{CompareCFT2} and
Fig.\ref{T20_DifferentLA}.
In particular, for $l=L/2$, $S_A(t)$ can be further simplified as
\be
S_A(t)\simeq \frac{c}{6}\log\frac{L}{\pi \epsilon}+\frac{c}{6}
\log\left[2-\cos\left(\frac{\sqrt{3}\pi}{L}t\right)\right],
\ee
which is Eq.\eqref{SA_oscillation00b} in the main text.

\subsubsection{Comparison with a single-quench }

As mentioned in the main text, in the high-frequency driving limit, one can consider the approximation
$e^{-H_0T}e^{-H_1T}\simeq e^{-(H_0+H_1)T}$, and then the Floquet dynamics can
be effectively described by a single quench with the effective Hamiltonian
$H_F=\frac{1}{2}(H_0+H_1)$. Here let us check this approximation explicitly.

In Ref.\onlinecite{WenWu2018}, we have considered a single quench starting from the ground state of
$H_0$, and switch the Hamiltonian to $H_{\text{M$\ddot{o}$b}}$ suddenly, with
\be
H_{\text{M$\ddot{o}$b}}(\theta)=H_0-\frac{\tanh(2\theta)}{2}(H_++H_-),
\ee
where
\be
\left\{
\begin{split}
H_0=&\int_0^L \frac{dx}{2\pi}T_{\tau\tau}(x)=\int_0^L\frac{dx}{2\pi}\left(T(w)+\bar{T}(\bar{w})\right),\\
H_{\pm}=&\int_0^L\frac{dx}{2\pi}
\left(
e^{\pm 2\pi w/L}T(w)+e^{\mp2\pi\bar{w}/L}\bar{T}(\bar{w})
\right).
\end{split}
\right.
\ee
Then the entanglement entropy evolution has the form: \cite{WenWu2018}
\be\label{SA_SignleQuench}
S_A(t)\simeq \frac{c}{6}\log\frac{L}{\pi \epsilon}+\frac{c}{12}\log\frac{f(t)^2+f(t)\cdot h(t)}{2}
\ee
where
\be
f(t)=\sqrt{h(t)^2+\sin^2\frac{2\pi l}{L}},
\ee
and
\be
\begin{split}
h(t)=&-\left(
\sin^2\frac{\pi t}{L_{\text{eff}}}\cdot \cosh(4\theta)+\cos^2\frac{\pi t}{L_{\text{eff}}}
\right)\cos\frac{2\pi l}{L}\\
&+\sin^2\frac{\pi t}{L_{\text{eff}}}\cdot \sinh(4\theta),
\end{split}
\ee
with $L_{\text{eff}}=L\cosh(2\theta)$.
Now we consider the high frequency limit of the Floquet CFT with $T_0=T_1=T$. Then one has
$H_F=\frac{1}{2}(H_0+H_1)=H_{\text{M$\ddot{o}$b}}(\theta)$ with
\be
\tanh(2\theta)=\frac{1}{2}.
\ee
Then one has $e^{4\theta}=3$, $\cosh(2\theta)=2/\sqrt{3}$, and $\sinh(2\theta)=1/\sqrt{3}$.
It is straightforward to check that the entanglement entropy evolution in Eq.\eqref{SA_SignleQuench}
is the same as the high-frequency limit of a Floquet CFT in Eqs.\eqref{SA_nonheating_SM} and \eqref{EFP}.

\subsubsection{Near the phase transition}
\label{NearPhaseTransition_NonHeating_SM}

\begin{figure}[tp]
\includegraphics[width=3.1in]{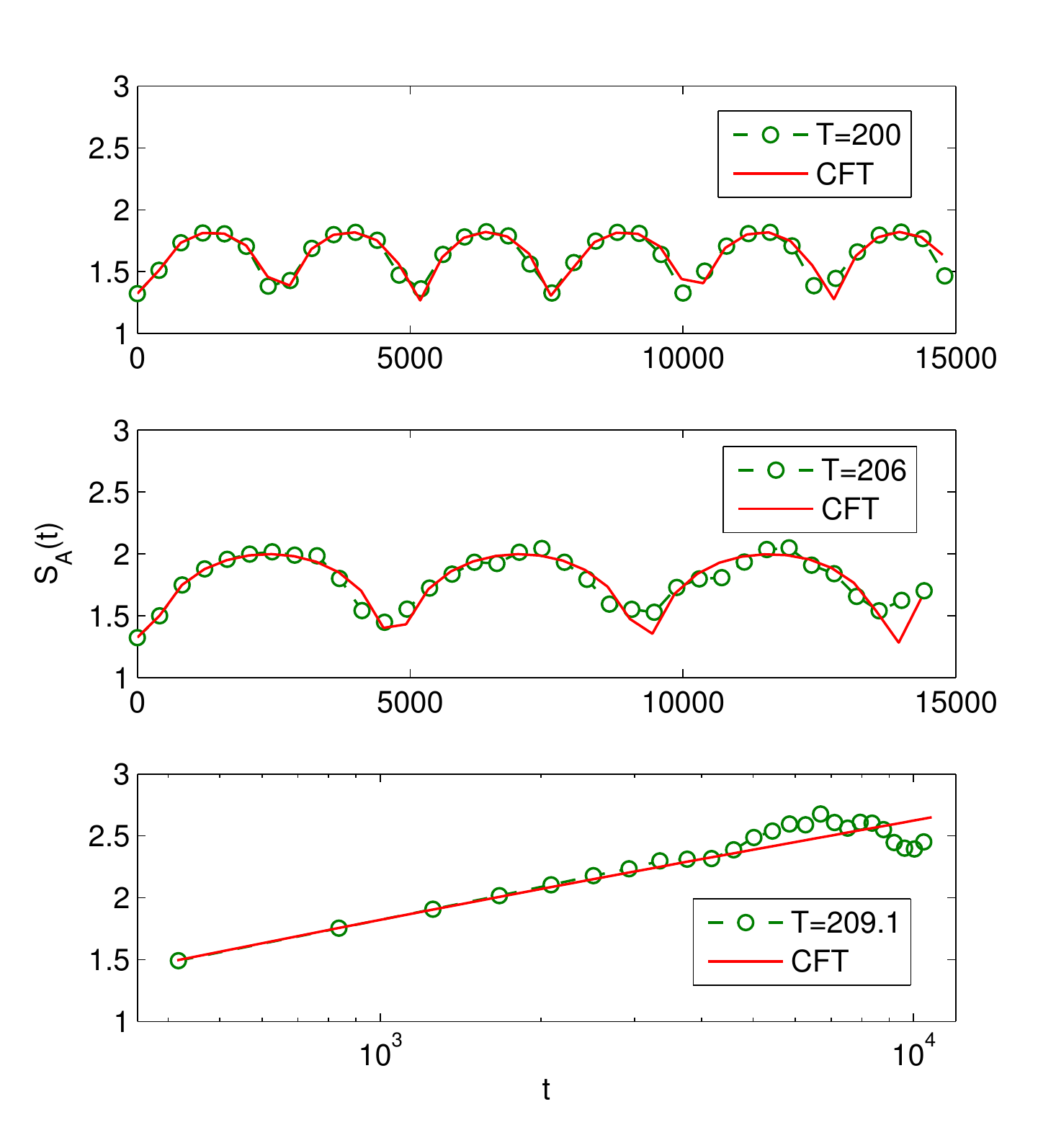}
\caption{
Comparison between numerics and CFT calculation for
entanglement evolution in the non-heating phase near the phase transition.
The driving period in CFT calculation is $T_0=T_1=T=199.5,\,205.1$, and $207.94$, respectively.
Here we choose $L=500$.
}
\label{CompareNearCritical}
\end{figure}

As we approach the phase transition from the side of non-heating phase, one can find that
the oscillation period of entanglement entropy diverges, as shown in Fig.\ref{OscillationCriticalPoint} in the main text.
[See also Fig.\ref{CompareNearCritical}.]

Now let us check the behavior of $T_E$ in Eq.\eqref{TE_osc_SM},
\textit{i.e.}, $T_E=\pi (T_0+T_1)/|\phi|$, near the phase transition explicitly.
Since there are two sets of solutions for the phase transition [see the main text],
here we consider them separately. We approach the phase transition along
\be
T_1=k \cdot T_0, \quad\text{for arbitrary } k>0.
\ee

One set of solution are $T_0=mL$, with $ m=1, 2, 3,\cdots$
[see the vertical lines in Fig.\ref{PhaseDiagram}].
Let us take
\be\label{PhaseTransition1_SM}
T_0^{\ast}=mL, \quad T_1^{\ast}=k T_0^{\ast}.
\ee
Since we approach the phase transition from the non-heating phase, then we make
$T_0=T_0^{\ast}+\delta$ and $T_1=T_1^{\ast}+k\delta$, with $\delta\ll T_0^{\ast}$.
Expanding to the first order in $\delta$, one has
\be
\Delta\simeq \frac{2\pi^2 km}{L} \delta.
\ee
After some straightforward algebra, one can obtain
\be
|\phi|\simeq \sqrt{\Delta} \simeq \kappa \delta^{1/2}=\kappa \cdot (T_0-T_0^{\ast})^{1/2},
\ee
where $\kappa:=\sqrt{2\pi^2km/L}$.
Therefore, near the phase transition in Eq.\eqref{PhaseTransition1_SM},
the oscillation period $T_E$ depends on $(T_0-T_0^{\ast})$ as follows
\be
T_E\simeq\frac{(1+k)mL}{\sqrt{2km/L}}\cdot \frac{1}{(T_0-T_0^{\ast})^{1/2}}.
\ee
In particular, for $T_0=T_1=T$, \textit{i.e.}, $k=1$, one has
\be
T_E\simeq \sqrt{2m}\cdot L^{3/2}\cdot \frac{1}{(T-T^{\ast})^{1/2}}.
\ee

The other set of solutions for the phase transition are determined by
$\Big[1-\left(\frac{\pi T_1}{L}\right)^2\Big]\sin\frac{\pi T_0}{L}+
2\cdot \frac{\pi T_1}{L}\cdot \cos\frac{\pi T_0}{L}=0$.
By choosing $(n-1)L<T_0^{\ast}<nL$,
$T_0=T_0^{\ast}-\delta$,
 and $T_1=k\cdot T_0$, one can find that
\be
\Delta\simeq \kappa'\cdot \delta+\mathcal{O}(\delta^2),
\ee
where $\kappa'=\sin\frac{\pi T_0^{\ast}}{L}
\Big\{
2\cdot\frac{\pi^2 T_1^{\ast}}{L^2}\cdot \sin\frac{\pi T_0^{\ast}}{L}\cdot (1+k)
-\frac{\pi}{L}\cdot \cos\frac{\pi T_0^{\ast}}{L}\cdot\big[1+2k-\big(\frac{\pi T_1^{\ast}}{L}\big)^2\big]
\Big\}.
$
Then one can obtain
\begin{small}
\be
T_E\simeq \Big|\frac{\pi T_1^{\ast}}{L}\sin\frac{\pi T_0^{\ast}}{L}-\cos\frac{\pi T_0^{\ast}}{L}\Big|
\cdot \frac{\pi (T_0^{\ast}+T_1^{\ast})}{\sqrt{\kappa'}}\cdot \frac{1}{(T_0^{\ast}-T_0)^{1/2}}.
\ee
\end{small}
That is, $T_E\propto 1/(T_0^{\ast}-T_0)^{1/2}$.

In a short summary, for the non-heating phase near the phase transitions,
one always has $\Delta\propto \delta$, and
$|\phi|\propto \delta^{1/2}$, based on which we can obtain
$T_E\propto |T_0-T_0^{\ast}|^{-1/2}$.

\subsection{Heating phase }

\subsubsection{Entanglement entropy evolution}

The heating phase corresponds to $\Delta<0$ in Eq.\eqref{Delta}.
After doing analytical continuation, one can find
\be\label{SAHeatingPhase_SM}
S_A(t)\simeq\frac{c}{6}\log\frac{L}{\pi \epsilon}
+\frac{c}{12}\,\log\frac{\mathcal{E}^2+\mathcal{F}^2-\mathcal{E}\sqrt{\mathcal{E}^2+\mathcal{F}^2}}{2P^4},
\ee
with $t=n(T_0+T_1)$ and
\be\label{EF_heating_SM}
\left\{
\begin{split}
\mathcal{E}=&-\cosh(2n\phi')\cdot\left(W \cdot \cos\frac{\pi T_0}{L}-\cos\frac{2\pi l}{L}\right)\\
&+\sinh(2n\phi')\cdot \left(P\cdot \sin\frac{\pi T_0}{L}\right)+K,\\
\mathcal{F}=&P^2\sin\frac{2\pi l}{L},
\end{split}
\right.
\ee
where
\be\label{Define_Heating_SM}
\left\{
\begin{split}
e^{2\phi'}:=&\frac{Q+P}{Q-P},\\
Q:=&\sin\frac{\pi T_0}{L}-\frac{L}{\pi T_1}\cos\frac{\pi T_0}{L},\\
P:=&\frac{L}{\pi T_1}\sqrt{-\Delta},\\
W:=&\cos\frac{\pi T_0}{L}+\frac{L}{\pi T_1}\sin\frac{\pi T_0}{L},\\
K:=&W\cdot \cos\frac{\pi T_0}{L}-W^2\cdot\cos\frac{2\pi l}{L}.
\end{split}
\right.
\ee
It is helpful to compare the parameters above with those in Eq.\eqref{Parameters_SM} for the non-heating phase.
As a self-consistent check, now let us look at the case with $n=0$, \textit{i.e.}, $t=0$.
Then one has $
\mathcal{E}=(1-W^2)\cos\frac{2\pi l}{L}=P^2\cos\frac{2\pi l}{L},
$
where we have used the fact that $W^2+P^2=1$.
Then $S_A(t)$ in Eq.\eqref{SAHeatingPhase_SM} can be simplified as
\be
S_A(t=0)\simeq \frac{c}{6}\log \left[\frac{L}{\pi\epsilon}\sin\left(\frac{\pi l}{L}\right)\right],
\ee
which is the entanglement entropy in the ground state, as expected.

\begin{figure}[tp]
\includegraphics[width=3.0in]{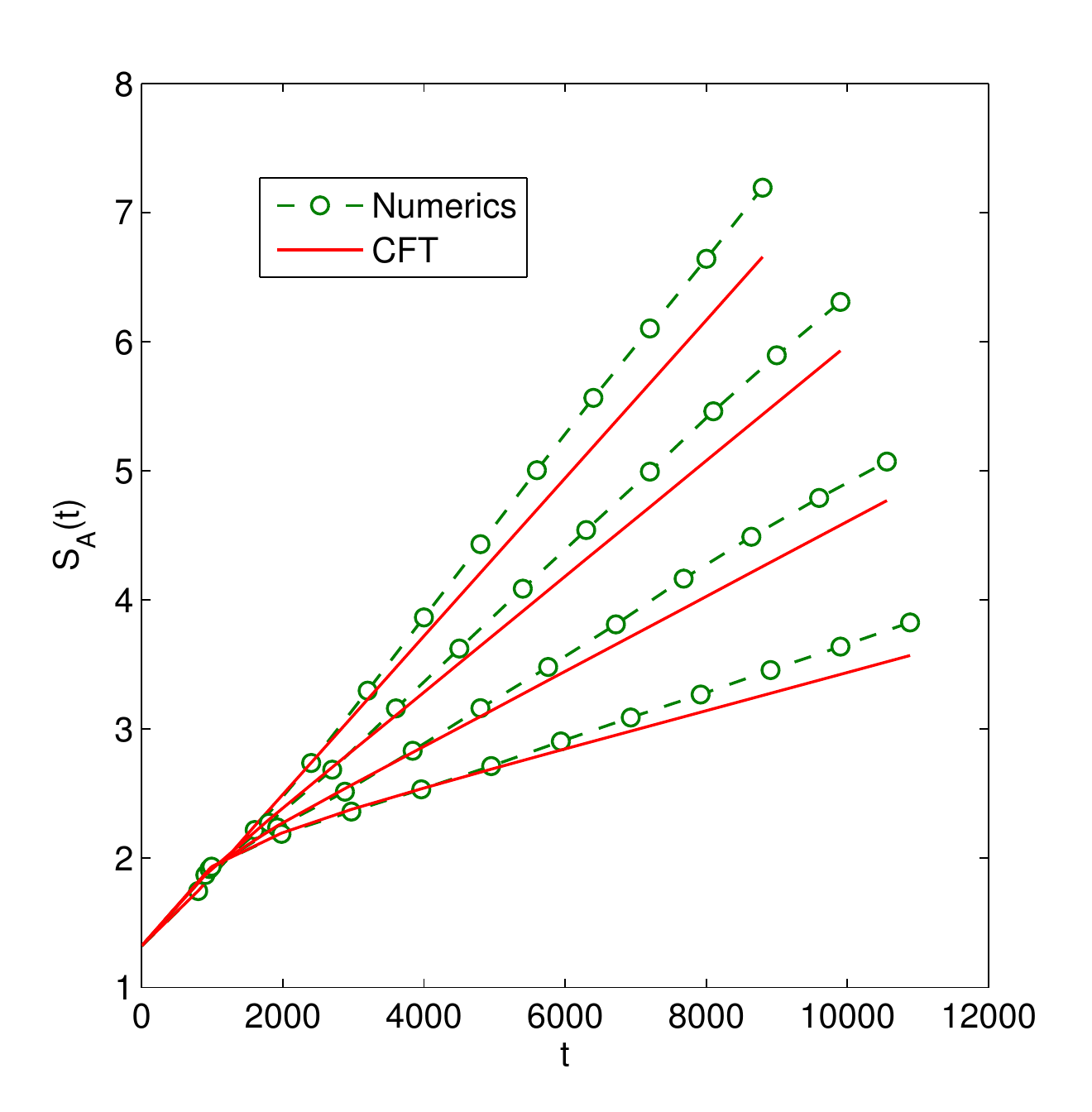}
\caption{
Comparison between numerics and CFT calculation for
entanglement evolution in the heating phase.
From top to bottom: $T_0=T_1=T=400,\,450,\,480$, and $495$.
Here we choose $L=500$.
}
\label{CompareHeatingPhase}
\end{figure}

Now let us check the specific case with $l=L/2$, so that $S_A(t)$ in Eq.\eqref{SAHeatingPhase_SM}
can be further simplified.
One can find that $\mathcal{F}=0$, $K=W\cdot \cos\frac{\pi T_0}{L}+W^2$, and
\be
\mathcal{E}=-R'\cosh(2n\phi'+\varphi')+K,
\ee
where we have defined
\be
\left\{
\begin{split}
R'e^{\varphi'}=&W\cos\frac{\pi T_0}{L}+1-P\sin\frac{\pi T_0}{L},\\
R'e^{-\varphi'}=&W\cos\frac{\pi T_0}{L}+1+P\sin\frac{\pi T_0}{L}.
\end{split}
\right.
\ee
Then the entanglement entropy can be expressed as
\be\label{SA_heating_SM}
S_A(t)\simeq \frac{c}{6}\log\Big[
\frac{L}{\pi \epsilon}\cdot\frac{R'\cosh(2n\phi'+\varphi')-K}{P^2}\Big].
\ee
At $n=0$, one can check that
\be
R'\cosh\varphi'-K=1-W^2=P^2.
\ee
Therefore, one has $S_A(t=0)=\frac{c}{6}\log\frac{L}{\pi\epsilon}$, which
is the entanglement entropy in the ground state. For generic $t$, one has
\be\label{SA_heating_SM2}
S_A(t)\simeq \frac{c}{6}\log\Big[
\frac{L}{\pi \epsilon}\cdot\frac{R'\cosh(2n\phi'+\varphi')-K}{R'\cosh\varphi'-K}\Big],
\ee
which is Eq.\eqref{SA_heating} in the main text.
A typical plot of $S_A(t)$ for different driving periods is shown in Fig.\ref{CompareHeatingPhase}.
It is noted that as $n$ grows, $S_A(t)$ grows linearly in time as follows
\be
S_A(t)\simeq \frac{c}{6}\log \frac{L}{\pi \epsilon}+\frac{c}{3}\cdot |\phi'|\cdot n.
\ee
Noting that $t:=n(T_0+T_1)$, one has
\be
S_A(t)\simeq \frac{c}{6}\log \frac{L}{\pi \epsilon}+\frac{c}{3}\cdot \frac{|\phi'|}{T_0+T_1}\cdot t, \quad
t=n(T_0+T_1).
\ee
In the above result, the entanglement entropy keeps growing linearly in time without saturation.
This is because in the conformal field theory, the energy spectrum goes to infinity without an upper bound,
and there are infinite degrees of freedom.
On a lattice, however, we always have a finite number of degrees of freedom
for a finite subsystem and the bandwidth of energy spectrum is finite.
Therefore the entanglement entropy will finally saturate, as will be discussed shortly.
(See Fig.\ref{LongTimeLimit})

Now let us check the entanglement entropy for an arbitrary subsystem $A=[0,l]$ with $0<l<L$.
Based on Eqs.\eqref{SAHeatingPhase_SM} and \eqref{EF_heating_SM}, it looks that for a generic $l$
the entanglement entropy will always grow linearly in time in the large $n$ limit.
However, this is not the case. Let us rewrite $\mathcal{E}$ in Eq.\eqref{EF_heating_SM} as follows
\be
\begin{split}
\mathcal{E}=&\frac{e^{2n\phi'}}{2}\left(-W\cos\frac{\pi T_0}{L}+\cos\frac{2\pi l}{L}+P\sin\frac{\pi T_0}{L}\right)\\
&+\frac{e^{-2n\phi'}}{2}\left(-W\cos\frac{\pi T_0}{L}+\cos\frac{2\pi l}{L}-P\sin\frac{\pi T_0}{L}\right)+K.\\
\end{split}
\ee
The entanglement entropy will grow linearly in time for large $n$ when satisfying:
(i) $\phi'>0$ and $-W\cos\frac{\pi T_0}{L}+\cos\frac{2\pi l}{L}+P\sin\frac{\pi T_0}{L}<0$,
or (ii) $\phi'<0$ and $-W\cos\frac{\pi T_0}{L}+\cos\frac{2\pi l}{L}-P\sin\frac{\pi T_0}{L}<0$.
One can check that for $l=L/2$, one of the above situations must be satisfied,
and therefore the entanglement entropy will always grow linearly in time for large $n$.

\begin{figure}[htp]
\includegraphics[width=3.0in]{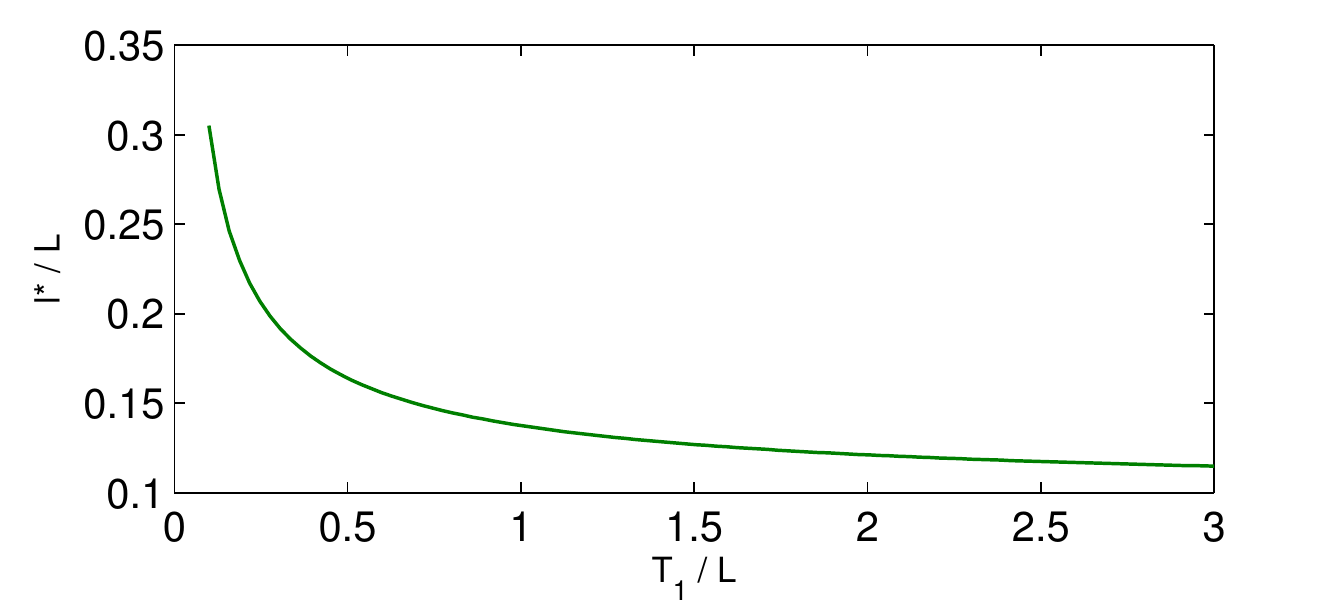}
\caption{
A typical plot of $l^{\ast}$ as a function of driving period $T_1$ in the heating phase.
Here we choose $T_0/L=0.9$.
}
\label{Lcritical}
\end{figure}

However, one can find there exists a length $l^{\ast}$, so that
for $l<l^{\ast}$, neither condition (i) nor (ii) is satisfied.
That is, for $A=[0,l]$ or $A=[L-l,L]$ with $l<l^{\ast}$, the entanglement entropy will not grow linearly
in time even for large $n$. In other words, the region with $l<l^{\ast}$ is not `heated', and only
the region in $(l^{\ast},L-l^{\ast})$ is `heated'. A typical plot of $l^{\ast}$ in the heating phase is
shown in Fig.\ref{Lcritical}.
For $l<l^{\ast}$, one can find that as $n$ grows, $S_A(t)$ will not grow linearly in time, but
evolve to a stable value with
\be\label{SA_stable}
S_A(n\phi'\gg 1)\simeq \frac{c}{6}\log \left(\frac{L}{2\pi \epsilon}\cdot \sin\frac{2\pi l}{L}\right),
\quad l<l^{\ast}.
\ee
As a remark, this is a typical feature in a quantum quench by quenching the ground state of $H_0$
[see Eq.\eqref{H0_maintext}] with a new Hamiltonian
$H_1=H_0-\frac{h}{2}\left(H_++H_-\right)$ where $h>1$ and
$H_{\pm}=\int_0^L\frac{dx}{2\pi}
\left(e^{\pm 2\pi w/L}T(w)+e^{\mp2\pi\bar{w}/L}\overline{T}(\bar{w})\right)$.
More details will be presented in \onlinecite{WenUN}.

In the lattice model under a periodic driving, we did not observe this stable behavior in Eq.\eqref{SA_stable}.
For arbitrary $l<L$ in a lattice model, we always observed a linear growth in $S_A(t)$ before
saturation [see Fig.\ref{LongTimeLimit}
for example].
This disagreement may result from the lattice effect, which we leave as a future problem.

\subsubsection{Near the phase transition}

As shown in Fig.\ref{SlopeChange}, the slope $k_E$ of linear growth of the entanglement entropy will vanish
near the phase transitions. In other words, $1/k_E$ diverges near the phase transitions.
In the following, we will show that as we approach the phase transition along $T_1=k\cdot T_0$, where
$k$ is an arbitrary positive real number, $1/k_E$ always diverges as
$1/k_E\propto |T_0-T_0^{\ast}|^{-1/2}$. The critical exponent $\zeta=1/2$ is the same as that
obtained from the side of non-heating phase.

The analysis is similar to that in the non-heating phase in Sec.\ref{NearPhaseTransition_NonHeating_SM}.
There are two sets of solutions for the phase transitions and let us discuss them separately.
First, for $T_0=mL$, with $m=1,\,2,\,3\cdots$,
[see the vertical lines in Fig.\ref{PhaseDiagram}].
Let us take
$T_0^{\ast}=mL$, $T_1^{\ast}=k T_0^{\ast}$.
Since we approach the phase transition from the heating phase, then we make
$T_0=T_0^{\ast}-\delta$, and $T_1=T_1^{\ast}-k\delta$.
Expanding to the first order in $\delta$, one has
\be
\Delta\simeq -\frac{2\pi^2 km}{L} \delta.
\ee
Based on the definitions in Eq.\eqref{Define_Heating_SM}, one can obtain
\be
|\phi'|\simeq \sqrt{-\Delta} \simeq \kappa \delta^{1/2}=\kappa \cdot (T_0^{\ast}-T_0)^{1/2},
\ee
where $\kappa:=\sqrt{2\pi^2km/L}$.
Therefore, near the phase transitions at $T_1=mL$ with $m=1,\,2,\,3\cdots$,
$1/k_E$ for the linear growth of entanglement entropy depends on
$(T_0-T_0^{\ast})$ as follows
\be
\frac{1}{k_E}=\frac{3(T_0+T_1)}{c}\cdot \frac{1}{|\phi'|}\simeq
\frac{3(T_0^{\ast}+T_1^{\ast})}{c\sqrt{2\pi^2km/L}}\cdot \frac{1}{(T_0^{\ast}-T_0)^{1/2}}.
\ee

The other set of solutions for the phase transition are determined by
$\Big[1-\left(\frac{\pi T_1}{L}\right)^2\Big]\sin\frac{\pi T_0}{L}+
2\cdot \frac{\pi T_1}{L}\cdot \cos\frac{\pi T_0}{L}=0$.
By choosing $(n-1)L<T_0^{\ast}<nL$,
$T_0=T_0^{\ast}+\delta$,
 and $T_1=k\cdot T_0$, one can find that
\be
\Delta\simeq -\kappa'\cdot \delta,
\ee
where $\kappa'=\sin\frac{\pi T_0^{\ast}}{L}
\Big\{
2\cdot\frac{\pi^2 T_1^{\ast}}{L^2}\cdot \sin\frac{\pi T_0^{\ast}}{L}\cdot (1+k)
-\frac{\pi}{L}\cdot \cos\frac{\pi T_0^{\ast}}{L}\cdot\big[1+2k-\big(\frac{\pi T_1^{\ast}}{L}\big)^2\big]
\Big\}.
$
Then one can obtain
\begin{small}
\be
\frac{1}{k_E}\simeq \Big|\frac{\pi T_1^{\ast}}{L}\sin\frac{\pi T_0^{\ast}}{L}-\cos\frac{\pi T_0^{\ast}}{L}\Big|
\cdot \frac{3 (T_0^{\ast}+T_1^{\ast})}{c\cdot \sqrt{\kappa'}}\cdot \frac{1}{(T_0-T_0^{\ast})^{1/2}}.
\ee
\end{small}
In short, for the heating phase near the phase transitions,
one always has $\Delta\propto -\delta$ and
$|\phi'|\propto \delta^{1/2}$, based on which we can obtain
$1/k_E\propto |T_0-T_0^{\ast}|^{-1/2}$.

As a short summary, by approaching the phase transitions from both the non-heating phase and the
heating phase, one can obtain the critical exponent $\zeta=1/2$ from the entanglement entropy evolution.

\subsubsection{Long time limit in a lattice model}

As seen from Eqs.\eqref{SA_heating} and \eqref{LinearGrowth} in the main text, the entanglement entropy
for $A=[0, L/2]$ grows linearly in time all the way, without saturation.
As we already mentioned, this is because there are infinite number of degrees of freedom inside the subsystem $A$
and the energy spectrum goes to infinity without an upper bound, so that the system can absorb energy all the way.
In a lattice model, however, the degrees of freedom in a finite subsystem are finite.
The bandwidth of energy spectrum is also finite.
It is expected that the entanglement entropy will saturate in the long time limit.

\begin{figure}[htp]
\includegraphics[width=2.9in]{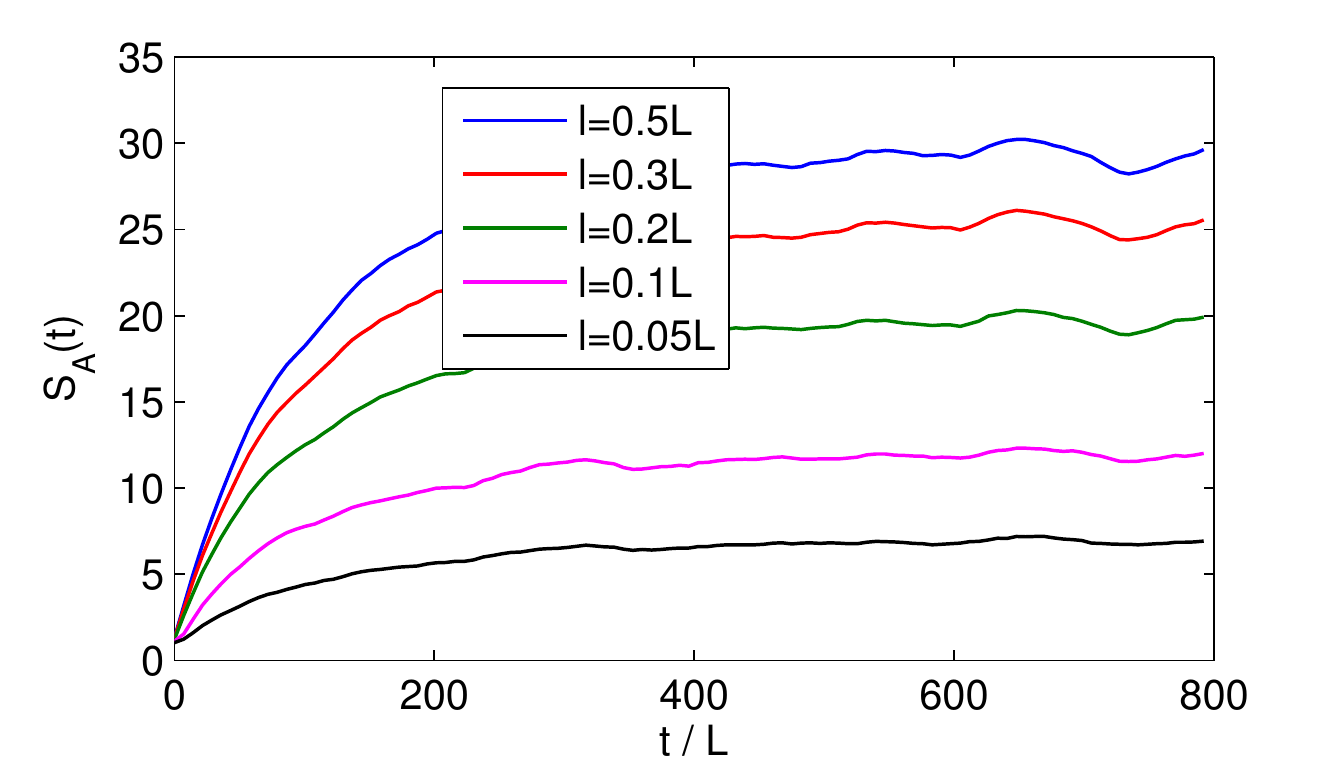}
\caption{
Numerical simulation for the entanglement entropy evolution in the long time limit in the heating phase.
We choose $T_1=T_2=T=0.9L$, with $L=500$.
}
\label{LongTimeLimit}
\end{figure}

As shown in Fig.\ref{LongTimeLimit}, we calculate the entanglement evolution in the long time limit on a free
fermion lattice (See Sec.\ref{Sec:Lattice} for the lattice model.).
The entanglement entropy grows linearly in time first, and then saturates, as expected.

At the current stage, in the field theory approach,
it is an open question for us to introduce the saturation in the entanglement evolution
in the heating phase.
(Note that this is different from the case of a global quench in CFTs where the saturation in
entanglement evolution is introduced
by the finite energy density in the initial state.\cite{CC_Global}
In the Floquet CFT, since we drive the system periodically, the system can absorb energy all the way
if there are infinite degrees of freedom and the energy spectrum goes to infinity.)
Similar problems
also appear in the entanglement entropy in a CFT with finite temperature.
In the high temperature limit, the entanglement entropy for a finite subsystem of length $l$ is
$S_A(\beta)\simeq \frac{c}{3}\log \frac{l}{\epsilon}+\frac{c}{3}\cdot \frac{\pi l}{\beta}$,
 where $\epsilon$ is the UV
cutoff introduced at the entanglement cut.\cite{cardy2016entanglement}
The entanglement entropy grows linearly with the temperature $1/\beta$ all the way. In a lattice system,
however, the entanglement entropy will finally saturate as $1/\beta$ increases, since
there are a finite number of degrees of freedom in a finite subsystem and the bandwidth
(of the energy spectrum) is finite.

\subsection{Phase transitions}

The phase transitions happen at $\Delta=0$ [see Eq.\eqref{Delta}].
To study the entanglement entropy at the phase transition, we cannot use the formula in Eq.\eqref{SA002}
directly. It is because after analytical continuation, one has $\eta=1$ and $\gamma_1=\gamma_2$. Therefore,
$\mathfrak{a}=\mathfrak{b}=\mathfrak{c}=\mathfrak{d}=0$ in Eq.\eqref{abcd_Ncycle},
and then Eq.\eqref{SA002} is not well defined.
In this case, $z_n$ is related to $z$ in Eq.\eqref{Zn_parabolic}, \textit{i.e.},
\be\label{Zn_parabolic_SM}
\frac{1}{z_n-\gamma}=\frac{1}{z-\gamma}+n\cdot \beta.
\ee
where $\gamma=(a-d)/2c$ and $\beta=c$, with $a,\,b,\,c,\,d$ given in Eq.\eqref{abcd_SM}.
Then Eq.\eqref{Zn_parabolic_SM} can be rewritten as
\be
z_n=\frac{\mathfrak{a}z+\mathfrak{b}}{\mathfrak{c}z+\mathfrak{d}}.
\ee
where
$\mathfrak{a}=1+n\beta\cdot \gamma$, $\mathfrak{b}=-n\beta\cdot \gamma^2$,
$\mathfrak{c}=n\beta$, and $\mathfrak{d}=1-n\beta\cdot \gamma$.
Then, following the procedure in Sec.\ref{EE_expression}, one can obtain the entanglement entropy at the phase
transitions as follows:
\be\label{PhaseTransitionEE_SM}
S_A(n)=\frac{c}{6}\log\frac{L}{\pi\epsilon}+\frac{c}{12}\log
\frac{\mathcal{E}^2+\mathcal{F}^2-\mathcal{E}\sqrt{\mathcal{E}^2+\mathcal{F}^2}}{2}.
\ee
Note that there are two sets of solutions for $\Delta=0$ at the phase transitions, and the expressions of
$\mathcal{E}$ and $\mathcal{F}$ are different for these two sets of solutions, as discussed in
the following.

\subsubsection{Phase transition I}

One set of solutions for the phase transitions are $T_0=mL$, with $m=1,\,2,\,3\cdots$,
which correspond to the vertical lines in Fig.\ref{PhaseDiagram}.
Here we denote this set of solutions as phase transition $I$.
In this case, one can find that the entanglement entropy has the expression in
Eq.\eqref{PhaseTransitionEE_SM} with
\be\label{PhaseTransitionI_EF}
\left\{
\begin{split}
\mathcal{E}=&-4n^2\left(\frac{\pi T_1}{L}\right)^2\cdot \sin^2\frac{\pi l}{L}+\cos\frac{2\pi l}{L},\\
\mathcal{F}=&\sin\frac{2\pi l}{L}.
\end{split}
\right.
\ee
One can check that for $t=0$, \textit{i.e.}, $n=0$, one has
\be
S_A(t=0)=\frac{c}{6}\cdot \log\left(
\frac{L}{\pi \epsilon}\cdot \sin\frac{\pi l}{L}
\right),
\ee
which is the entanglement entropy in the ground state, as expected.
For $l=L/2$, the entanglement entropy can be simplified as
\be\label{PhaseTransitionEE01_SM}
S_A(n)=\frac{c}{6}\log \frac{L}{\pi \epsilon}+\frac{c}{6}\log\left[
1+4n^2\cdot\left(\frac{\pi T_1}{L}\right)^2
\right],
\ee
which is Eq.\eqref{SAlogMiddle01} in the main text. A typical plot is shown in Fig.\ref{PhaseTransitionIandII} (top).

As a remark, it is interesting that $S_A(n)$ at phase transition $I$ has the same form
as that after a single quench in Ref.\onlinecite{WenWu2018}. In Ref.\onlinecite{WenWu2018},
we start from the ground state of $H_0$, and evolve it with the new Hamiltonian $H_1$.
Then the entanglement entropy evolution has the expression in Eq.\eqref{PhaseTransitionEE_SM} with
\be\label{EF_SingleQuench}
\left\{
\begin{split}
\mathcal{E}=&-4\left(\frac{\pi t}{L}\right)^2\sin^2\frac{\pi l}{L}+\cos\frac{2\pi l}{L},\\
\mathcal{F}=&\sin\frac{2\pi l}{L}.
\end{split}
\right.
\ee
By making $t=nT_1$, Eqs.\eqref{PhaseTransitionI_EF} and \eqref{EF_SingleQuench} are the same.
This is not a coincidence.
In the case of Floquet CFTs, for $T_0=mL$ with $m=1,\,2,\,3\cdots$, the state `revives'
after a time evolution of $T_0$ with $H_0$.
Effectively, the state only evolves according to $H_1$, corresponding to the single-quench case.
This can be easily seen based on Eq.\eqref{z1}. After one cycle of driving, one has
(after analytical continuation)
\be
z_1=\frac{
\left[(1+\frac{i\pi T_1}{L})\cdot e^{\frac{i\pi T_0}{L}}\right]z
-\frac{i\pi T_1}{L}\cdot e^{\frac{-i\pi T_0}{L}}
}{
\left(\frac{i\pi T_1}{L}\cdot e^{\frac{i\pi T_0}{L}}\right)z
+\left(1-\frac{i\pi T_1}{L}\right)\cdot e^{\frac{-i\pi T_0}{L}}
}.
\ee
One can find that for $T_0=mL$ with $m=1,\,2,\,3\cdots$,
$z_1$ has the same form as that for $T_0=0$.
I.e., effectively, the state only evolves with the Hamiltonian $H_1$.

\subsubsection{Phase transition II}

\begin{figure}[tp]
\includegraphics[width=3.1in]{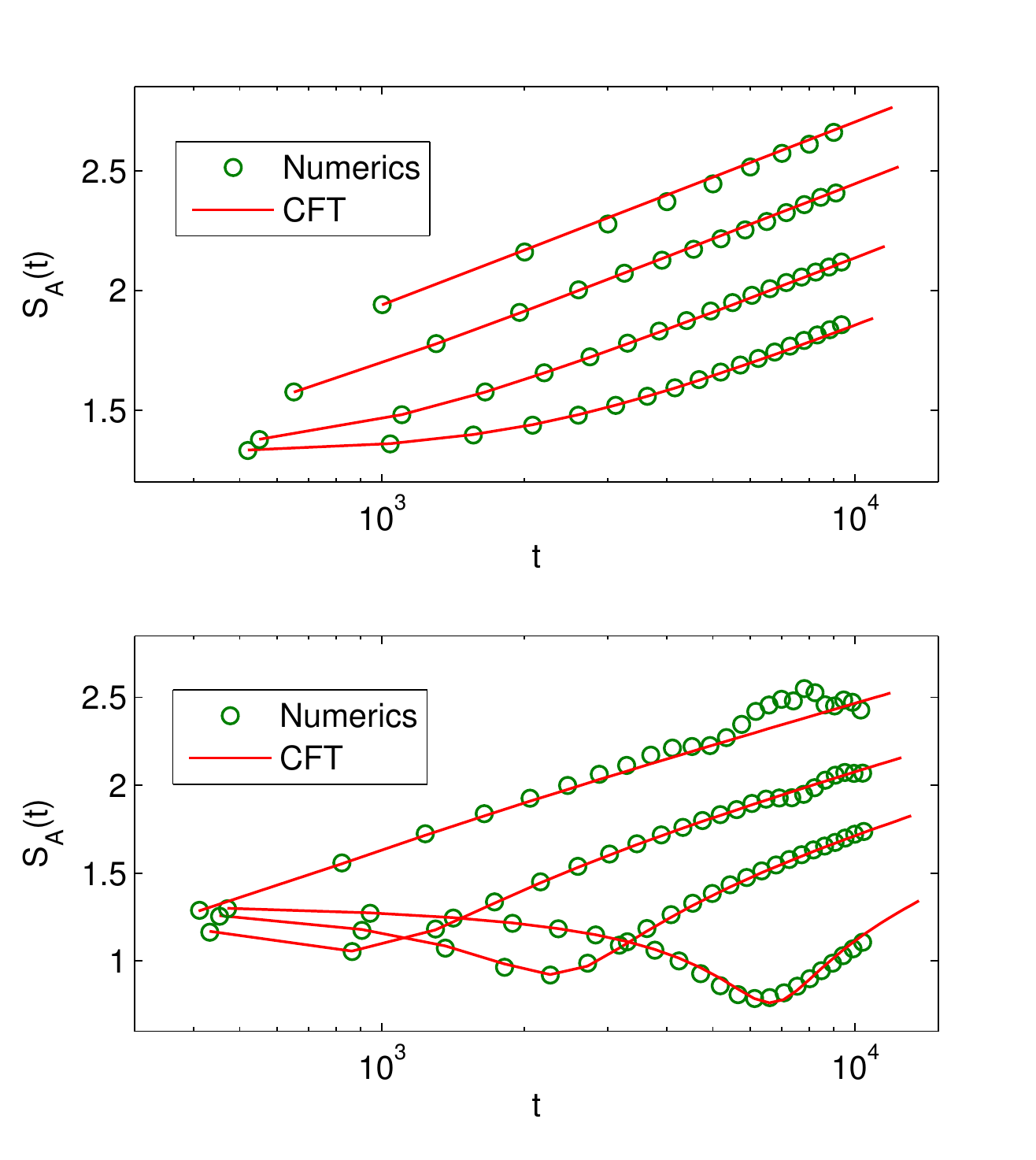}
\caption{
Entanglement entropy evolution at the phase transition.
(Top)
Phase transition at $T_0=L$, with $T_1=500,\,150,\,50,\,$ and $20$ (from top to bottom);
and (Bottom) phase transition for $T_0<L$ with $T_1=150,\,80,\,50$ and $30$ (from top to bottom).
Here we choose $L=500$, and $l=L/2$.
}
\label{PhaseTransitionIandII}
\end{figure}

The other set of solutions for phase transitions are determined by
$\Big[1-\left(\frac{\pi T_1}{L}\right)^2\Big]\sin\frac{\pi T_0}{L}+
2\cdot \frac{\pi T_1}{L}\cdot \cos\frac{\pi T_0}{L}=0$. We denote
this set of solution as phase transition $II$.
It can be found that the entanglement entropy has the expression in
Eq.\eqref{PhaseTransitionEE_SM} but with $\mathcal{E}$ and $\mathcal{F}$ as follows
\be
\left\{
\begin{split}
\mathcal{E}=&-\Big[
n^2\cdot\left(\frac{2(1-x^2)}{x^2(1+x^2)}-\frac{2}{x^2}\cos\frac{2\pi l}{L}\right)\\
&-n\cdot\frac{4}{1+x^2}-\cos\frac{2\pi l}{L}
\Big]\\
\mathcal{F}=&\sin\frac{2\pi l}{L}.
\end{split}
\right.
\ee
where we have defined
$x:=\frac{L}{\pi T_1}$.
As a self-consistent check, for $n=0$, one has
$
\mathcal{E}=\cos\frac{2\pi l}{L}.
$
Then one can find that
\be
S_A=\frac{c}{6}\log\left(
\frac{L}{\pi \epsilon}\cdot \sin\frac{\pi l}{L}
\right),
\ee
which is the entanglement entropy in the ground state, as expected.
For the specific case with $l=L/2$, one has
\be
\mathcal{E}=-\left[
n^2\cdot \frac{4}{x^2(1+x^2)}-n\cdot \frac{4}{1+x^2}+1
\right],
\ee
and $\mathcal{F}=0$. Then the entanglement entropy can be simplified as
\be
S_A(t)\simeq \frac{c}{6}\log \Bigg\{\frac{L}{\pi \epsilon}\Big(
\frac{4 \left(\frac{\pi T_1}{L}\right)^4}{1+\left(\frac{\pi T_1}{L}\right)^2}\, n^2
-\frac{4\left(\frac{\pi T_1}{L}\right)^2}{1+\left(\frac{\pi T_1}{L}\right)^2}\, n+1
\Big)\Bigg\}.\nonumber
\ee
which is Eq.\eqref{SAlogSolution2} in the main text. A typical plot of $S_A(t)$ is shown
in Fig.\ref{PhaseTransitionIandII} (bottom).
It is noted that for both phase transitions $I$ and $II$, the entanglement entropy grows
as $S_A(t)\simeq \frac{c}{3}\log t$ for large $n$.

\subsection{On single-point correlation function}

Since the entanglement entropy in this work is calculated through the correlation function
of twist operators which are themselves primary operators, it is straightforward to obtain the
correlation functions from the results of entanglement entropy (and vice versa).
This can be clearly seen in Eqs.\eqref{Oww002}$\sim$\eqref{SA002}.

As an example, for a primary operator in the non-heating phase, one can find that
\be
\begin{split}
&\langle\psi(t)|\mathcal{O}(x=l)|\psi(t)\rangle\\
=&A_{\mathcal{O}}^b\cdot\left(\frac{\pi\epsilon}{L}\right)^{2h}\cdot\left(
\frac{2P^4}{\mathcal{E}^2+\mathcal{F}^2-\mathcal{E}\sqrt{\mathcal{E}^2+\mathcal{F}^2}}
\right)^h
\end{split}
\ee
where $h$ is the conformal dimension, and $P$, $\mathcal{E}$, and $\mathcal{F}$
are given in Eqs.\eqref{EF_nonheating0A_SM} and \eqref{Parameters_SM}.
For $l=L/2$, one has
\be
\begin{split}
&\langle\psi(t)|\mathcal{O}(x=L/2)|\psi(t)\rangle\\
=&A_{\mathcal{O}}^b\cdot
\left(\frac{\pi \epsilon}{L}
\cdot
\frac{K-R\cos\varphi}{K-R\cos(2n\phi+\varphi)}
\right)^{2h}.
\end{split}
\ee
See Eq.\eqref{Parameters_SM} for definitions of variables.
That is, the single-point correlation function oscillates in time with $t:=n(T_0+T_1)$.
Similarly, one can find that the single-point correlation function decays exponentially in time
in the heating phase, and decays in a power-law in time at the phase transitions.

The behaviors of entanglement entropy and single-point correlation functions and their
correspondence with M$\ddot{\text{o}}$bius transformations are summarized in Table I.

\section{A lattice model on critical fermion chain}
\label{Sec:Lattice}

Here we give some details on the calculation of entanglement entropy in a free fermion lattice
under periodic driving. The essential part is to calculate the equal time two-point correlation functions.
Then based on the method in \onlinecite{peschel2003calculation}, one can evaluate the
entanglement entropy explicitly.

We consider a free fermion chain with half filling. It has finite sites $L$ with open
boundary conditions.
The Hamiltonians $H_0$ and $H_1$ have the following form:\cite{WenWu2018}
\be
\left\{
\begin{split}
H_0=&\frac{1}{2}\sum_{i=1}^{L-1}c_i^{\dag}c_{i+1}+h.c.,\\
H_1=&\sum_{i=1}^{L-1}\sin^2\left(\frac{\pi (i+1/2)}{L}\right)c_i^{\dag}c_{i+1}+h.c.
\end{split}
\right.
\ee
where $c_i$ ($c_i^{\dag}$) are fermionic operators, which satisfy the
anticommutation relations $\{c_i,c_j\}=\{c_i^{\dag},c_j^{\dag}\}=0$, and
$\{c_i,c_j^{\dag}\}=\delta_{ij}$.
At $t=0$, we prepare the initial state as the ground state $|G\rangle$ of $H_0$, and then
evolve the state with $H_1$ for time $T_1$, and $H_0$ for time $T_0$.
Then we repeat this driving procedure in time.

For completeness, in the following we list the procedures of calculating two-point
correlation functions in various cases.

\subsection{Ground state}

Now we consider the ground state $|G\rangle$ of $H_0$, and evaluate
$\langle G|c_m^{\dag}c_n|G\rangle$.
With a unitary transformation
\be
c_n=\sum_{i} U_{ni}\gamma_i,\quad \gamma_i=\sum_j (U^{\dag})_{ij} c_j,
\ee
one can diagonalize the Hamiltonian $H_0$ as follows
\be
H_0=\sum_{i=1}^L \epsilon_i \gamma_i^{\dag}\gamma_i.
\ee
Note also that
$c_n^{\dag}=\sum_{i}\gamma_i^{\dag}U^{\ast}_{ni}=\sum_{i}\gamma_i^{\dag}(U^{\dag})_{in}$, and
$\gamma_i^{\dag}=\sum_j (U^{\dag})_{ij}^{\ast} c_j^{\dag}=\sum_j c_j^{\dag}U_{ji}$.
The ground state $|G\rangle$ can be written as
\be
|G\rangle=\prod_{i=1}^{L/2}\gamma_i^{\dag}|\text{vac}\rangle.
\ee
Then one can find
\be
\begin{split}
\langle G|c_m^{\dag}c_n|G\rangle
=& \sum_{i\in \text{occ.}} U_{ni} (U^{\dag})_{im},
\end{split}
\ee
where `$\text{occ.}$' denote the occupied modes.

\subsection{Quantum quench and Floquet case}

Now we consider a single quench by evolving the ground state $|G\rangle$ with the Hamiltonian $H_1$.
Then the time dependent wavefunction has the form
\be
|\psi(t)\rangle=e^{-iH_1t}|G\rangle.
\ee
Now we evaluate $\langle \psi(t)|c_m^{\dag}c_n|\psi(t)\rangle$.
We need another unitary transformation to diagonalize $H_1$, \textit{i.e.},
\be
c_n=\sum_{i} V_{ni}\beta_i,\quad \beta_i=\sum_j (V^{\dag})_{ij} c_j,
\ee
so that
\be
H_1=\sum_i \epsilon^1_i \beta_i^{\dag}\beta_i.
\ee
Note also that
$c_n^{\dag}=\sum_{i}\beta_i^{\dag}V^{\ast}_{ni}=\sum_{i}\beta_i^{\dag}(V^{\dag})_{in}$,
$\beta_i^{\dag}=\sum_j (V^{\dag})_{ij}^{\ast} c_j^{\dag}=\sum_j c_j^{\dag}V_{ji}$,
$e^{iH_1t}\,\beta_j^{\dag}\,e^{-iH_1t}=e^{i\epsilon^1_j t}\beta_j^{\dag}$, and
$e^{iH_1t}\,\beta_j \,e^{-iH_1t}=e^{-i\epsilon^1_{j} t}\beta_j$.
Then one can find that
\be
\begin{split}
\beta_i=&\sum_j (V^{\dag})_{ij} c_j =  \sum_j (V^{\dag})_{ij}\sum_k U_{jk}\gamma_k\\
=&\sum_k(V^{\dag}U)_{ik}\gamma_k
=:\sum_k W_{ik}\gamma_k.
\end{split}
\ee
That is,
\be
\beta_i=\sum_j W_{ij}\gamma_j.
\ee
where we have defined
$W=V^{\dag}U$.
Similarly, one has
\be
\gamma_k=\sum_i(W^{\dag})_{ki}\beta_i.
\ee
Then we can check that
\be
e^{iH_1t}c_ne^{-iH_1t}|G\rangle
=\sum_iV_{ni}e^{-i\epsilon^1_{i}t}\sum_k W_{ik}\gamma_k |G\rangle.
\ee
It is convenient to define
\be
W_{ik}^{(t),1}:=e^{-i\epsilon_i^1t}W_{ik},
\ee
and then
\be
\begin{split}
e^{iH_1t}c_ne^{-iH_1t}|G\rangle
=&\sum_{i}V_{ni}\sum_kW_{ik}^{(t),1}\gamma_k|G\rangle\\
=&\sum_i\big[
V\cdot W^{(t),1}
\big]_{nk}\gamma_k|G\rangle.
\end{split}
\ee
Then, it is straightforward to check that
\be
\begin{split}
&\langle G|e^{iH_1t}c_m^{\dag}e^{-iH_1t}e^{iH_1t}c_ne^{-iH_1t}|G\rangle\\
=&\langle G|\gamma_{k'}^{\dag}\big[V\cdot W^{(t),1}\big]^{\dag}_{k'm}\sum_k[V\cdot W^{(t),1}]_{nk}\gamma_k|G\rangle\\
=&\sum_{k\in \text{occ.}}
[V\cdot W^{(t),1}]_{nk}\cdot [V\cdot W^{(t),1}\big]^{\dag}_{km}.
\end{split}
\ee
Let us move one step further to the `double quench',
and consider the state $|\psi(t)\rangle=e^{-iH_0T_0}e^{-iH_1T_1}|G\rangle$, with
$t=T_0+T_1$.
We check the following quantity:
\be
\begin{split}
&e^{iH_1T_1}e^{iH_0T_0}c_ne^{-iH_0T_0}e^{-iH_1T_1}|G\rangle\\
=&\sum_i U_{ni}e^{-i\epsilon^0_iT_0} \sum_j(W^{\dag})_{ij}e^{-i\epsilon^1_j T_1}\sum_k W_{jk}\gamma_k|G\rangle.
\end{split}
\ee
By defining
\be
W_{ik}^{(T_0),0}=e^{-i\epsilon_i^0T_0}\big(W\big)^{\dag}_{ik},
\ee
 one has
\be
\begin{split}
&e^{iH_1T_1}e^{iH_0T_0}c_ne^{-iH_0T_0}e^{-iH_1T_1}|G\rangle\\
=&\sum_{k\in \text{occ.}} \Big[U\cdot W^{(T_0),0}\cdot W^{(T_1),1}\Big]_{nk}\gamma_k|G\rangle.
\end{split}
\ee
Then it is straightforward to obtain
\be
\begin{split}
\langle \psi(t)|c_m^{\dag}c_n|\psi(t)\rangle
=&
\sum_{k\in \text{occ.}}
\Big[U\cdot W^{(T_0),0}\cdot W^{(T_1),1}\Big]_{nk}\cdot \\
&\Big[U\cdot W^{(T_0),0}\cdot W^{(T_1),1}\Big]^{\dag}_{km}.
\end{split}
\ee
Now we consider the Floquet case, with
\be
|\psi(t)\rangle=e^{-iH_0T_0}e^{-iH_1T_1}\cdots e^{-iH_0T_0}e^{-iH_1T_1}|G\rangle,
\ee
where $t=n(T_0+T_1)$.
Based on the above examples, it is straightforward to obtain
\be
\langle \psi(t)|c_m^{\dag}c_n|\psi(t)\rangle=\sum_{k\in\text{occ.}}
\mathcal{W}_{nk}\cdot (\mathcal{W}^{\dag})_{km},
\ee
where
\be
\mathcal{W}=U\cdot [W^{(T_0),0}\cdot W^{(T_1),1}]\cdots [W^{(T_0),0}\cdot W^{(T_1),1}].
\ee
In the above, we showed how to obtain the two-point correlation functions for various cases,
based on which we can obtain the entanglement entropy evolution by using the method in
Ref.\onlinecite{peschel2003calculation}.

\end{document}